%% file: main.tex
\documentclass[a4paper, 11pt]{article}

\usepackage{jheppub}
\usepackage{booktabs}
\usepackage[noabbrev]{cleveref}
\usepackage[acronym,shortcuts]{glossaries-extra}
\setabbreviationstyle[acronym]{long-short}
\glsdisablehyper
\usepackage[italic]{hepnames}
\usepackage{physics}
\usepackage{siunitx}
\usepackage{subcaption}
\usepackage{todonotes}

\newacronym{ape}{APE}{Array Processor Experiment}
\newacronym{apbc}{aPBC}{anti-periodic boundary conditions}
\newacronym{bc}{BC}{boundary condition}
\newacronym{cg}{CG}{Clebsch-Gordan}
\newacronym{cls}{CLS}{Coordinated Lattice Simulations}
\newacronym{cm}{CM}{centre-of-mass}
\newacronym{chpt}{ChPT}{chiral perturbation theory}
\newacronym{dcsb}{DCSB}{dynamical chiral symmetry breaking}
\newacronym{ddhmc}{DD-HMC}{domain decomposition HMC}
\newacronym{ddvcs}{DDVCS}{double deeply virtual Compton scattering}
\newacronym{dis}{DIS}{deep inelastic scattering}
\newacronym{dr}{DR}{dimensional regularisation}
\newacronym{ds}{DS}{Dyson-Schwinger}
\newacronym{dvcs}{DVCS}{deeply virtual Compton scattering}
\newacronym{dvmp}{DVMP}{deeply virtual meson production}
\newacronym{eic}{EIC}{Electron-Ion Collider}
\newacronym{eicc}{EicC}{Electron-ion collider in China}
\newacronym{erbl}{ERBL}{Efremov-Radyushkin-Brodsky-Lepage}
\newacronym{fse}{FSE}{finite-size effect}
\newacronym{gcr}{GCR}{generalized conjugate residual}
\newacronym{gevp}{GEVP}{Generalized Eigenvalue Problem}
\newacronym{hep}{HEP}{high-energy physics}
\newacronym{hmc}{HMC}{hybrid Monte Carlo}
\newacronym{ib}{IB}{isospin breaking}
\newacronym{ir}{IR}{infrared}
\newacronym{irrep}{irrep}{irreducible representation}
\newacronym{ldme}{LDME}{long-distance matrix element}
\newacronym{lhc}{LHC}{Large Hadron Collider}
\newacronym{lhs}{lhs}{left-hand-side}
\newacronym{lo}{LO}{leading order}
\newacronym{lqcd}{LQCD}{lattice quantum chromodynamics}
\newacronym{mphmc}{MP-HMC}{mass preconditioned HMC}
\newacronym{nlo}{NLO}{next-to-leading order}
\newacronym{nnlo}{NNLO}{next-to-next-to-leading order}
\newacronym{n3lo}{N3LO}{next-to-next-to-next-to-leading order}
\newacronym{nrqcd}{NRQCD}{non-relativistic QCD}
\newacronym{ope}{OPE}{operator product expansion}
\newacronym{ozi}{OZI}{Okubo–Zweig–Iizuka}
\newacronym{pbc}{PBC}{periodic boundary condition}
\newacronym{ptbc}{PTBC}{partially twisted boundary conditions}
\newacronym{qcd}{QCD}{quantum chromodynamics}
\newacronym{qed}{QED}{quantum electrodynamics}
\newacronym{qft}{QFT}{quantum field theory}
\newacronym{rgi}{RGI}{renormalization group invariant}
\newacronym{rhs}{rhs}{right-hand side}
\newacronym{sap}{SAP}{Schwarz alternating procedure}
\newacronym{sdf}{SDF}{short distance factorization}
\newacronym{slac}{SLAC}{Stanford Linear Accelerator Center}
\newacronym{svd}{SVD}{singular value decomposition}
\newacronym{tbc}{TBC}{twisted boundary condition}
\newacronym{tcs}{TCS}{timelike Compton scattering}
\newacronym{uv}{UV}{ultraviolet}
\newacronym{varpro}{VP}{variable projection}

\newacronym{gda}{GDA}{generalized distribution amplitude}
\newacronym{gpd}{GPD}{generalized parton distribution}
\newacronym{lcda}{LCDA}{light-cone DA}
\newacronym{da}{DA}{distribution amplitude}
\newacronym{pda}{pDA}{pseudo-DA}
\newacronym{pdf}{PDF}{parton distribution function}
\newacronym{ppdf}{pPDF}{pseudo-PDF}
\newacronym{itpdf}{ITPDF}{Ioffe-time PDF}
\newacronym{itd}{ITD}{Ioffe-time DA}
\newacronym{pitd}{pITD}{Ioffe-time pseudo-DA}
\newacronym{rpitd}{rpITD}{reduced Ioffe-time pseudo-DA}
\newacronym{lamet}{LaMET}{Large-momentum effecive theory}

\AtBeginDocument{\RenewCommandCopy\qty\SI}
\DeclareUnicodeCharacter{0302}{^}

\DeclareMathOperator{\Ci}{Ci}
\DeclareMathOperator{\Si}{Si}

\newcommand{\iu}{\text{i}}

\newcommand{\msbar}{\ensuremath{\overline{\text{MS}}}}
\newcommand{\tmat}{\ensuremath{T}}

\arxivnumber{2406.04668}
\title{\boldmath The distribution amplitude of the $\Petac$-meson at leading twist from Lattice QCD}

\author[a]{B. Blossier,}
\author[b]{M. Mangin-Brinet,}
\author[c]{J.M. Morgado Ch\'{a}vez}
\author[a]{and T. San Jos\'{e}}

\affiliation[a]{Laboratoire de Physique des 2 Infinis Irène Joliot-Curie, CNRS/IN2P3,\\
Université Paris-Saclay, 91405 Orsay Cedex, France}
\affiliation[b]{Laboratoire de Physique Subatomique et de Cosmologie, CNRS/IN2P3,\\
38026 Grenoble Cedex, France}
\affiliation[c]{D\'{e}partement de Physique Nucl\'{e}aire, Irfu/CEA-Saclay,\\
91191 Gif-sur-Yvette Cedex, France}

\emailAdd{blossier@ijclab.in2p3.fr}
\emailAdd{mariane@lpsc.in2p3.fr}
\emailAdd{jose-manuel.morgadochavez@cea.fr}
\emailAdd{san-jose-perez@ijclab.in2p3.fr}

\abstract{Distribution amplitudes are functions of non-perturbative matrix elements describing the hadronization of quarks and gluons. Thanks to factorization theorems, they can be used to compute the scattering amplitude of high-energy processes. Recently, new ideas have allowed their computation using lattice QCD, which should provide us with a general, fully relativistic determination. We present the first lattice calculation of the $\Petac$-meson distribution amplitude at leading twist. Starting from the relevant matrix element in discrete Euclidean space on a set of $N_f=2$ CLS ensembles, we explain the method to connect to continuum Minkowski spacetime. After addressing several sources of systematic uncertainty, we compare to Dyson-Schwinger and non-relativistic QCD determinations of this quantity. We find significant deviations between the latter and our result even at small Ioffe times.}

\begin{document}
\maketitle
\flushbottom

\input{sections/introduction}
\input{sections/pseudo-da}
\input{sections/lattice}
\input{sections/extrapolation}
\input{sections/systematics}
\input{sections/comparisons}

\input{sections/conclusions}

\acknowledgments

The work by J.M. Morgado Ch\'{a}vez has been supported by P2IO LabEx (ANR-10-LABX-0038) in the framework of Investissements d’Avenir (ANR-11-IDEX-0003-01). The work by T. San Jos\'{e} is supported by Agence Nationale de la Recherche under the contract ANR-17-CE31-0019. This project was granted access to the HPC resources of TGCC (2021-A0100502271, 2022-A0120502271 and 2023-A0140502271) by GENCI. The authors thank Michael Fucilla, Cédric Mezrag, Lech Szymanowski, and Samuel Wallon for valuable discussions.

\appendix

\input{appendices/varpro}
\input{appendices/moments}
\input{appendices/nrqcd}
\input{appendices/ds}

\bibliographystyle{JHEP}
\bibliography{biblio.bib}

\end{document}

%% file: sections/introduction.tex
\section{Introduction}

In 1964, Gell-Mann and Zweig postulated the existence of quarks as the underlying degrees of freedom in the eightfold way, the scheme to classify hadrons introduced back in 1961. Following their hypothesis, Taylor, Kendall and Friedman conducted a series of experiments in \glsxtrshort{slac} between 1966 and 1978, where they discovered the inner structure of hadrons. Their studies consisted of \gls{dis} reactions probing protons and neutrons with electrons at high energies. Ever since, different inclusive and exclusive processes have been discovered to offer a privileged window to the structure of baryons and mesons. We focus in those exclusive scattering processes where there is a large momentum transfer to the hadron target and factorization theorems hold \cite{Collins:1989gx,Radyushkin:1996ru,Collins:1996fb}. They allow to separate the amplitude in two main pieces: the scattering of the probe with quarks and gluons and the internal structure of the target. The first is calculable in perturbation theory and it is described by coefficient functions, while the second is a consequence of \glsxtrshort{qcd} confinement at low energies and it is given by a variety of soft distributions, like \glspl{gpd} and \glspl{da}. Examples of processes where the \gls{da} appears are meson photoproduction \cite{Lepage:1980fj}, $\HepProcess{\Pphoton^*\Pphoton^* \to M}$, and \gls{dvmp}, in which an off-shell photon interacts with a target, a proton for example, and a meson is produced in the final state $\HepProcess{\Pphoton^* \Pproton \to M \Pproton}$.

The \gls{da} of a single meson, our object of interest, is a particular case of the more universal \glspl{gda}, which consider interacting hadrons, and depend on further degrees of freedom. \Glspl{gda} and \glspl{gpd} are connected via the crossing symmetry, allowing for a more uniform theoretical study of these quantities, see \cite{Diehl:2003ny} for a review. Unfortunately, the non-perturbative nature of these functions makes them difficult to compute, and usually contribute the most to the cross-section uncertainty. In turn, this makes extracting information from experiments more difficult.
Traditionally, the \gls{da} has been studied using \gls{nrqcd} \cite{Bodwin:1994jh,Chung:2019ota}, \gls{ds} equations \cite{Ding:2015rkn,Binosi:2018rht,Serna:2020txe}, light-front dynamics \cite{Arifi:2024mff} or light-cone sum rules \cite{Braun:1997kw,Braun:2006dg}. Of especial interest for our work are the studies of charmonium \glspl{da} employing the latter method in
\cite{Braguta:2006wr,Braguta:2007fh,Braguta:2007tq,Braguta:2008qe}.
As these quantities are non-pertubative, it would be reasonable to think that lattice simulations are a natural choice to compute \glspl{gpd}, \glspl{da} and other functions from first principles. Several works exist that reconstruct the \gls{da} from its first moments \cite{RQCD:2019osh,RQCD:2019hps}. However, it was not until 2013 that fundamental limitations were overcome in the seminal paper by Ji \cite{Ji:2013dva}. At the beginning, many efforts focused on the \gls{pdf} of the nucleon \cite{Alexandrou:2019lfo,Karpie:2021pap} and the \gls{pdf} and \gls{da} of the pion and kaon \cite{Joo:2019bzr,Gao:2022vyh,LatticeParton:2022zqc,Baker:2024zcd,Kovner:2024pwl,Holligan:2024umc}, while other works aim now to compute their \glspl{gpd} \cite{Bhattacharya:2024qpp,Dutrieux:2024umu} as well as the structure of heavy mesons \cite{Zhao:2020bsx}. See \cite{Constantinou:2020pek} for a review on the lattice progress. The latter are especially important to benefit the most from the \glsxtrshort{eic}, \glsxtrshort{eicc} and \glsxtrshort{lhc} experiments, and with this project we aim at contributing to this effort. To this end, we present an \textit{ab initio} calculation of the $\Petac$-meson \gls{da} at leading twist, the first employing \glsxtrshort{lqcd}.

Let us start by defining the quantity of interest. The quark \gls{da} for a pseudoscalar state, which was introduced in 1977 \cite{Radyushkin:1977gp} for the particular case of the pion, is given by the Fourier transform of a bi-local matrix element. In particular, using the light-cone metric in the light-cone gauge, $A^+=0$, the \gls{da} is given by \cite{Diehl:2003ny}
\begin{equation}
\label{eq:definition-da}
\phi(x) = \int \dfrac{\dd{z^-}}{2\pi} e^{-\iu (x-1/2) p^+ z^-} \eval{\mel{\Petac(p)}{\APcharm(-z/2)\gamma^+\gamma_5\Pcharm(z/2)}{0}}_{z^+=z_{\text{T}}=0},
\end{equation}
where $\bra*{\Petac}$ is the pseudoscalar meson in the final state, $\ket{0}$ is the \acs{qcd} vacuum, and $\Pcharm$ and $\APcharm$ are the quark fields. The distance $z=(z^+,z^-,z_{\text{T}})$, with $z^+=z_{\text{T}}=0$, separates the quark fields and lies on the light-cone, $z^2=0$. Therefore, we may choose the direction $p^+$ for the $\Petac$ momentum. The definition in \cref{eq:definition-da} may include an additional $\iu$ factor to match the definition of the pseudoscalar decay constant, all while having a real-valued function $\phi$. However, this factor $\iu$ simply redefines the phase of the meson state and it is not observable.
Due to Lorentz invariance, this function depends solely on the plus-momentum fraction of the quark with respect to the meson, $x=q^+/p^+$. In other gauges, a straight Wilson line $W(-z^-/2,z^-/2)$ appears between the two field positions $-z^-/2$ and $z^-/2$ to ensure gauge invariance,
\begin{equation}
W(a,b) = P\exp\left(\iu g\int_b^a \dd{z^-} A^+(z^-n _-)\right)
\end{equation}
where $P$ indicates path ordering from $a$ to $b$. Neither \glspl{da}, nor \glspl{gpd} or other related functions can be computed using lattice simulations because, in Euclidean metric, only the null vector lies on the light cone. In \cite{Ji:2013dva}, Ji proposed a generalization of \glspl{pdf} to space-like separations, e.g. $z=(0,0,z_3,0)$ and $p=(0,0,p_3,E)$ in Euclidean space, commonly known as quasi-\glspl{pdf}, which tend towards the light-cone \glspl{pdf} in the infinite momentum limit $p_3\to\infty$. A similar approach can be applied to \glspl{da} and \glspl{gpd} \cite{Radyushkin:2019owq}. However, it was noted by Radyushkin in \cite{Radyushkin:2017cyf} that the dependence of quasi-\glspl{pdf} on their momentum is rather complicated, and that $p_3 \geq \qty{3}{\giga\eV}$ are necessary to approach their behavior at infinity. This requirement is still difficult to meet and, in particular, our set of ensembles have lattice spacings too coarse to reach the desired momentum.

In the same work \cite{Radyushkin:2017cyf}, an alternative method was proposed, known as pseudo distributions, which are related via Fourier transform to Ioffe-time pseudo-distributions. The latter are given by bi-local matrix elements in terms of the Ioffe time $\nu=pz=p_3 z_3$ \cite{Ioffe:1969kf} and $z_3^2$. These pseudo distributions also generalize the light-cone distributions to spacelike intervals, and tend towards them in the short distance limit $z_3^2 \to 0$. The factor $1/z_3^2$ plays an analogous role to a renormalization scale $\mu^2$. In our work, we employ the proposal in \cite{Radyushkin:2017cyf} as we understand the systematics are more favorable for our setup.


The remainder of this paper is organized as follows: In \cref{sec:pseudo-da-extraction}, we define the fundamental object which is computed on the lattice, the \gls{rpitd}, we explain how we extract it, how we model the \gls{lcda}, and how we relate the two. In \cref{sec:lattice-setup}, we detail our lattice calculation and the set of $N_f=2$ ensembles that we use. In \cref{sec:continuum-limit}, we take the continuum limit using three different lattice spacings. Since our ensembles include a variety of pion masses, we also take into account the subleading quark dependence. Finally, we present there the leading-twist \gls{da} of the $\Petac$-meson, which is the main result of this work. \Cref{sec:systematics} is devoted to estimate several sources of systematic uncertainties, while \cref{sec:comparisons} compares our results with alternative methods relying on \gls{nrqcd} and \gls{ds} equations. We give our conclusions in \cref{sec:conclusions}.


%% file: sections/pseudo-da.tex
\section{Methodology}
\label{sec:pseudo-da-extraction}

In Euclidean metric, we start from the matrix element \cite{Radyushkin:2016hsy}
\begin{equation}
    \label{eq:mel-lattice}
    M^{\alpha}(p,z) = e^{-\iu pz/2} \mel{\Petac(p)}{\APcharm(0)\gamma^{\alpha}\gamma_5 W(0,z)\Pcharm(z)}{0}
\end{equation}
where $W$ is the $0 \to z$ straight Wilson line, $\bra{\Petac(p)}$ is the pseudoscalar meson state with momentum $p$, $\ket{0}$ is the \glsxtrshort{qcd} vacuum, $\APcharm$ and $\Pcharm$ are quark fields, and the Wilson line $W$ along the vector $z$ ensures gauge invariance of the matrix element. Computing the matrix element in an asymmetric configuration as in \cref{eq:mel-lattice} allows to access all the lattice sites in the simulation, and to connect with the symmetric definition given in \cref{eq:definition-da}, we use translation invariance and multiply by the appropriate phase, $\exp(-\iu pz/2)$. A Lorentz decomposition divides $M^{\alpha}$ in two pieces \cite{Radyushkin:2016hsy},
\begin{equation}
    M^{\alpha}(p,z) = 2p^{\alpha}\mathcal{M}(p,z)+z^{\alpha}\mathcal{M}^{\prime}(p,z)
\end{equation}
where $\mathcal{M}$ carries both the leading-twist contribution and a higher-twist contamination at $\order{z^2\Lambda^2_{\text{QCD}}}$, while $\mathcal{M}^{\prime}$ is a purely higher-twist effect. To remove $\mathcal{M}^{\prime}$, we align the momentum along the z-axis, $p=(0,0,p_3,E)$, set an equal-time separation $z=(0,0,z_3,0)$ and select $\alpha=4$, such that $M^4(p,z)=2E \mathcal{M}(p,z)$. Note that $\mathcal{M}$ is a Lorentz invariant, and therefore it only depends on scalar combinations of $p$ and $z$, which are $\nu$ and $z^2$. To take the continuum limit the operator $\APcharm(0)\gamma^{\alpha}\gamma_5 W(0,z)\Pcharm(z)$ needs to be renormalized,
and while standard \glspl{da} $\phi(x,\mu)$ are defined in the $\msbar$ scheme and exhibit a logarithmic dependence on the renormalization scale $\mu$ \cite{Diehl:2003ny,Belitsky:2005qn},
\glspl{pda} $\phi(x,z_3^2)$ have ultraviolet singularities and diverge logarithmically in the limit $z^2 \to 0$.
To suppress the first, the authors of \cite{Radyushkin:2016hsy,Orginos:2017kos,Karpie:2018zaz} use the results from \cite{Ishikawa:2017faj}, where it is proven that the entire operator is multiplicatively renormalizable. Since the renormalization constant only depends on $z_3$, we cancel it forming a \gls{rgi} ratio that can be factorized using the $\msbar$ scheme into the \gls{da} and Wilson coefficients. The result is the \glsxtrshort{rgi} ratio \cite{Radyushkin:2016hsy,Orginos:2017kos,Karpie:2018zaz}
\begin{equation}
    \label{eq:rgi-ratio}
    \tilde{\phi}(\nu,z) \equiv \dfrac{\mathcal{M}(p,z) \mathcal{M}(0,0)}{\mathcal{M}(0,z) \mathcal{M}(p,0)}
\end{equation}
where $\nu=pz$ is the Ioffe time. The ratio $\tilde{\phi}(\nu,z)$ is a Lorentz scalar, and it depends solely on $\nu$ and $z^2$, but for notational convenience we suppress the square. We call \cref{eq:rgi-ratio} the \glsentryfull{rpitd} from now on, and this is the actual quantity extracted from the lattice data. Let us briefly discuss the form of \cref{eq:rgi-ratio}. The factor $\mathcal{M}(0,z)$ cancels the renormalization factor of $\mathcal{M}(p,z)$, and it corresponds to a local axial-vector current in the limit $z=0$. Said current should be conserved and normalized. However, this is not the case in lattice simulations due to lattice artifacts. To cancel the latter, we form the ratios $\mathcal{M}(p,z)/\mathcal{M}(p,0)$ and $\mathcal{M}(0,z)/\mathcal{M}(0,0)$, and then we divide the first by the second, cancelling the renormalization factors. This cancellation holds at all orders in perturbation theory \cite{Ishikawa:2017faj}, and the double ratio in \cref{eq:rgi-ratio} has a well defined continuum limit and does not require an additional renormalization factor.

Regarding the logarithmic divergences in the limit $z^2 \to 0$, the matching relation \cref{eq:pseudo-to-light-cone} cancels them. Specifically, upon removing the \glsxtrshort{qft} regulator and any remaining higher-twist contamination (see \cref{sec:continuum-limit}), we can relate the leading-
twist \gls{rpitd} $\tilde{\phi}_{\text{lt}}(\nu,z)$ and the light-cone \gls{itd} $\tilde{\phi}_{\text{lt}}(\nu,\mu)$ in the $\msbar$ scheme via the matching kernel $C$, which is derived in \cite{Radyushkin:2017lvu,Radyushkin:2019owq} at \gls{nlo} in perturbation theory. Since the \gls{rpitd} is \gls{rgi}, its $z^2$ behavior is scheme-independent, but its dependence on this scale must match the $\mu$ dependence of the $\msbar$ \gls{itd}. After performing the matching, we can Fourier-transform to obtain the \gls{lcda} at leading twist $\phi_{\text{lt}}(x,\mu)$,
\begin{equation}
    \label{eq:pseudo-to-light-cone}
    \tilde{\phi}_{\text{lt}}(\nu,z) = \int_0^1 \dd{w} C(w,\nu,z\mu) \int_0^1 \dd{x} \cos(wx\nu-w\nu/2) \phi_{\text{lt}}(x,\mu),
\end{equation}
where we choose the renormalization scale $\mu=\qty{3}{\giga\eV}$ throughout this work. Inverting the Fourier transform in \cref{eq:pseudo-to-light-cone} with only a limited number of data points in $\tilde{\phi}_{\text{lt}}(\nu,z)$ is an ill-posed problem which requires adding some extra information.
In particular, we parametrize the \gls{da} $\phi_{\text{lt}}(x,\mu)$ in terms of the shifted Gegenbauer polynomials $\tilde{G}(x) \equiv G(-1+2x)$ \cite[\href{https://dlmf.nist.gov/18}{chapter 18}]{NIST:DLMF}, where $x\in[0,1]$ and $G$ are the standard ultraspherical polynomials defined in the domain $g\in[-1,1]$. The left-hand side is determined thanks to the lattice simulations. Furthermore, we exchange the order of integration, expanding the cosine in a series of $\tilde{G}$ and computing the moments of $C$ in powers of $w$.

Let us explain more in detail the procedure. Starting with the \gls{lcda} parameterization, at leading twist and leading order in $\alpha_s$, the Gegenbauer polynomials $\tilde{G}^{(3/2)}(x)$ are eigenvectors of the \gls{erbl} equations describing the \gls{da} evolution \cite{Braun:2003rp}. This means that, at this order, one can express the \gls{da} as a polynomial series with coefficient $3/2$ \cite{Efremov:1979qk,Lepage:1979zb},
\begin{equation}
    \label{eq:conformal-expansion}
    \phi_{\text{lt}}(x,\mu) = 6x(1-x)\sum_{n=0}^{\infty} d^{(3/2)}_{n}(\mu) \tilde{G}_{n}^{(3/2)}(x).
\end{equation}
The \gls{da} is then an analytic function normalized to one, $\int\dd{x}\phi=1$. In this project, we assume that \cref{eq:conformal-expansion} is also a reasonable model of the \gls{da} that we compute non-perturbatively. If we were to use the infinite series of polynomials, the latter could describe functions like $1/x$ and $1/(1-x)$, and so the \gls{da} does not have to vanish at the endpoints $x=0$ and $x=1$. In practice, we truncate \cref{eq:conformal-expansion} at small $n$, and the \gls{da} vanishes at the boundaries. Since the lattice data provides no information in this region, we are introducing some model dependence.
After this discussion, we need to adapt \cref{eq:conformal-expansion} to the particular case of charmonium, where the \gls{da} should be symmetric around $x=1/2$. Since Gegenbauer polynomials with $n$ odd are anti-symmetric, their corresponding coefficients vanish. Besides, we can leave the coefficient $3/2$ undetermined and fit it to our data to speed up the convergence of the series. Changing the coefficient, which we call $\lambda$ from now on, amounts to choosing one particular basis of polynomials, but it is always possible to transform back to \cref{eq:conformal-expansion}. Therefore, we will employ the following expression to describe the charmonium light-cone \gls{da},
\begin{equation}
    \label{eq:light-cone-da-model}
    \phi_{\text{lt}}(x,\mu)
    = (1-x)^{\lambda-1/2}x^{\lambda-1/2} \sum_{n=0}^{\infty} d_{2n}^{(\lambda)} \tilde{G}_{2n}^{(\lambda)}(x),
    \qquad\qquad
    d_0^{(\lambda)} = \dfrac{4^{\lambda}}{B\left(1/2, \lambda+1/2\right)}
\end{equation}
where $B$ is a beta function and the cofficients $\lambda$ and $d_{2n}$ will be constrained with our lattice data. \Cref{eq:light-cone-da-model} is normalized to one and we can recover \cref{eq:conformal-expansion} replacing $\lambda=3/2$. A similar approach has been used in the context of \glsxtrshortpl{pdf} \cite{Karpie:2021pap}.

\begin{figure}[htbp]
    \centering
    \includegraphics{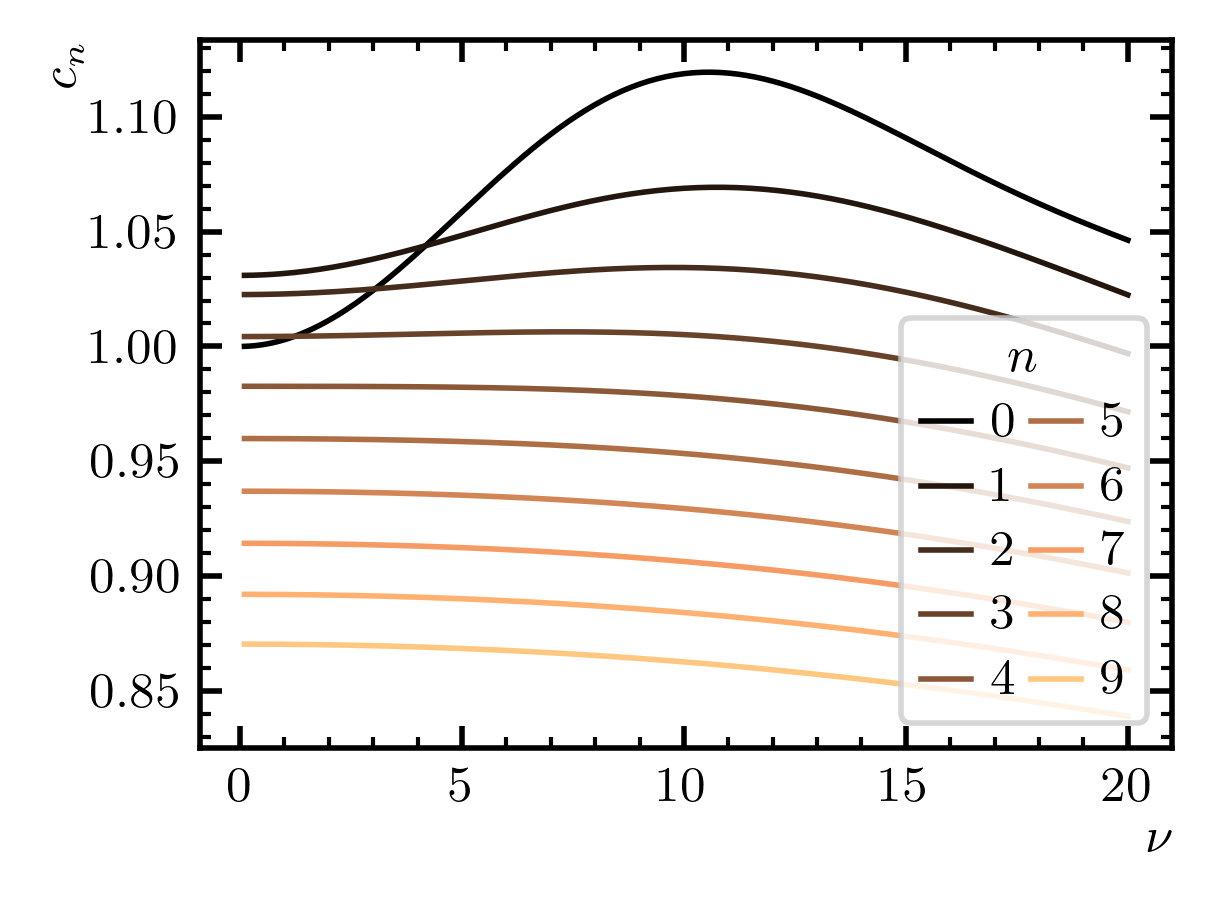}
    \caption{Moments $c_n(\nu, z_3\mu)$ of the \gls{da} matching kernel versus Ioffe time.
    For illustration, we set $z_3 = 5 \times \qty{0.0658}{\femto\meter}$.}
    \label{fig:cn}
\end{figure}

The next step is to compute the moments of the matching kernel $C$ derived in \cite{Radyushkin:2017lvu,Radyushkin:2019owq},
\begin{equation}
    \label{eq:cn-coefficient}
    c_n(\nu, z\mu)
    = \int_0^1 \dd{w} C(w,\nu,z\mu) w^n
    = 1-\dfrac{\alpha_{\text{s}}C_{\text{F}}}{2\pi}
    \left[\log\left(\dfrac{\mu^2}{\mu_0^2}\right)b_n(\nu)+l_n(\nu)\right]
\end{equation}
where $C_F=4/3$ is the Casimir in the fundamental representation, $\mu=\qty{3}{\giga\eV}$, and the initial energy scale $\mu_0$ is given by the Wilson line,
\begin{equation}
    \dfrac{1}{\mu_0^2} \equiv \dfrac{z^2~e^{2\gamma_{\text{E}}+1}}{4}
\end{equation}
where $\gamma_{\text{E}}$ is the Euler constant. We use the strong coupling constant in the $\overline{\textup{MS}}$-scheme \cite{Workman:2022ynf} $\alpha_s=0.2243$, together with $\Lambda \equiv \Lambda^{(2)}_{\text{QCD}} = \qty{330}{\mega\eV}$ \cite{FlavourLatticeAveragingGroupFLAG:2021npn} and $n_f=2$ flavors. The functions $b_n$ and $l_n$ are given in terms of hypergeometric functions ${}_pF_q$,
\begin{equation}
    \label{eq:bn-coefficient}
    \begin{aligned}
        b_n(\nu)
        &=-\dfrac{1}{2}-\sum_{j=0}^{n-1} \dfrac{2}{j+2}
        ~{}_1F_2\left(1,\dfrac{j+3}{2},\dfrac{j+4}{2},-\dfrac{\nu^2}{16}\right)
        -\dfrac{\nu^2}{24}~{}_2F_3\left(1,1,2,2,\dfrac{5}{2},-\dfrac{\nu^2}{16}\right)
        \\
        &+\dfrac{1}{(n+2)(n+1)}
        ~{}_1F_2\left(1,\dfrac{n+3}{2},\dfrac{n+4}{2},-\dfrac{\nu^2}{16}\right)
    \end{aligned}
\end{equation}
and
\begin{equation}
    \label{eq:ln-coefficient}
    \begin{aligned}
        l_n(\nu)
        &=1+4\sum_{j=0}^{n-1}
        \begin{pmatrix}
          n \\
          j+1
        \end{pmatrix}
        \dfrac{(-1)^j}{(j+1)^2}
        ~{}_2F_3\left(\dfrac{j+1}{2},\dfrac{j+1}{2},\dfrac{1}{2},\dfrac{j+3}{2},\dfrac{j+3}{2},-\dfrac{\nu^2}{16}\right)
        \\
        &+\dfrac{\nu^2}{8}~{}_3F_{4}\left(1,1,1,\dfrac{3}{2},2,2,2,-\dfrac{\nu^2}{16}\right)
        -\dfrac{2}{(n+2)(n+1)}~{}_1F_{2}\left(1,\dfrac{n+3}{2},\dfrac{n+4}{2},-\dfrac{\nu^2}{16}\right).
    \end{aligned}
\end{equation}
Note that all the hypergeometric functions ${}_pF_q$ in \cref{eq:bn-coefficient,eq:ln-coefficient} fulfill $p \leq q$, which is sufficient to prove that they converge for all values of Ioffe time \cite{196938}.
In \cref{fig:cn}, we plot $c_n$ for several $n$'s as a function of Ioffe time. In our analysis we will only use the lines for even $n$'s.

\begin{figure}
    \centering
    \begin{subfigure}[b]{0.49\textwidth}
        \centering
        \includegraphics{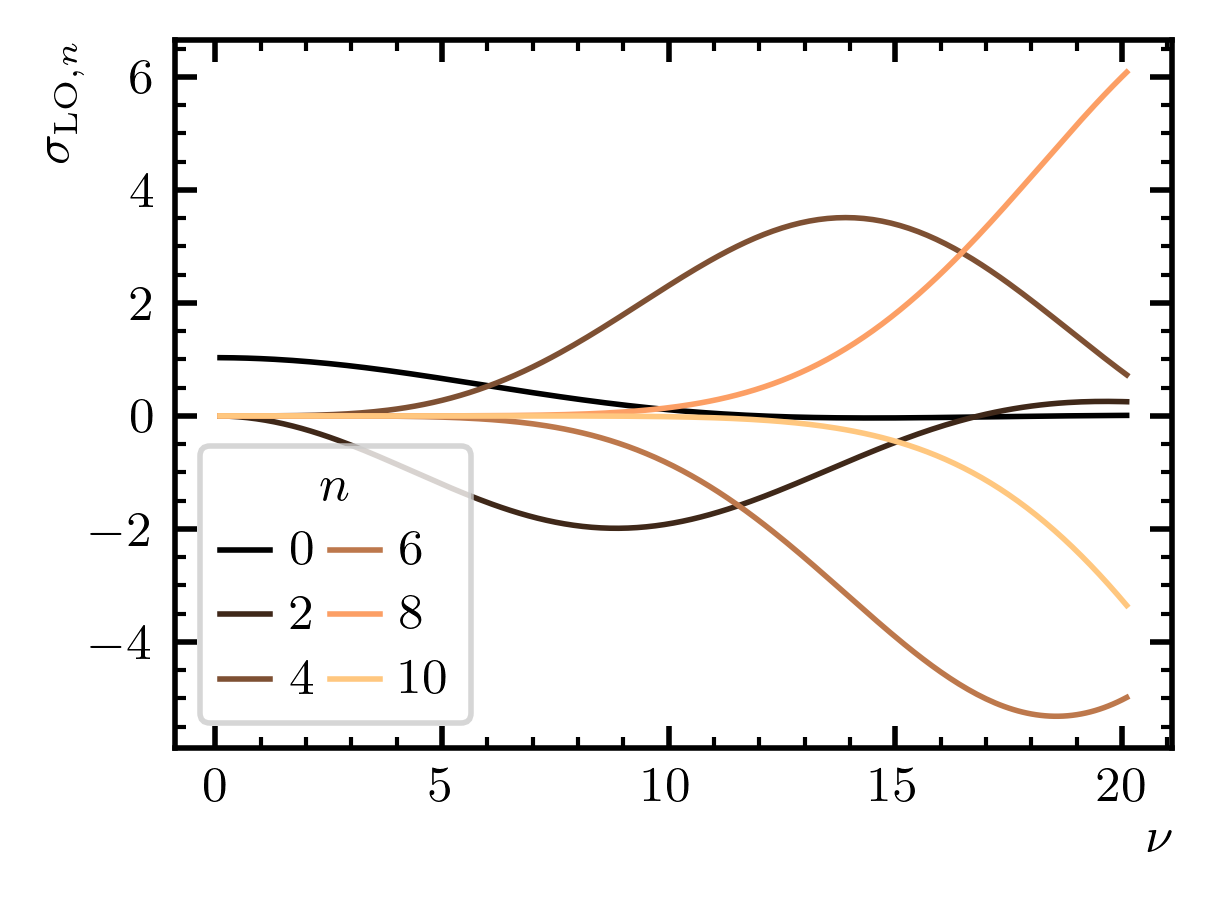}
    \end{subfigure}
    \begin{subfigure}[b]{0.49\textwidth}
        \centering
        \includegraphics{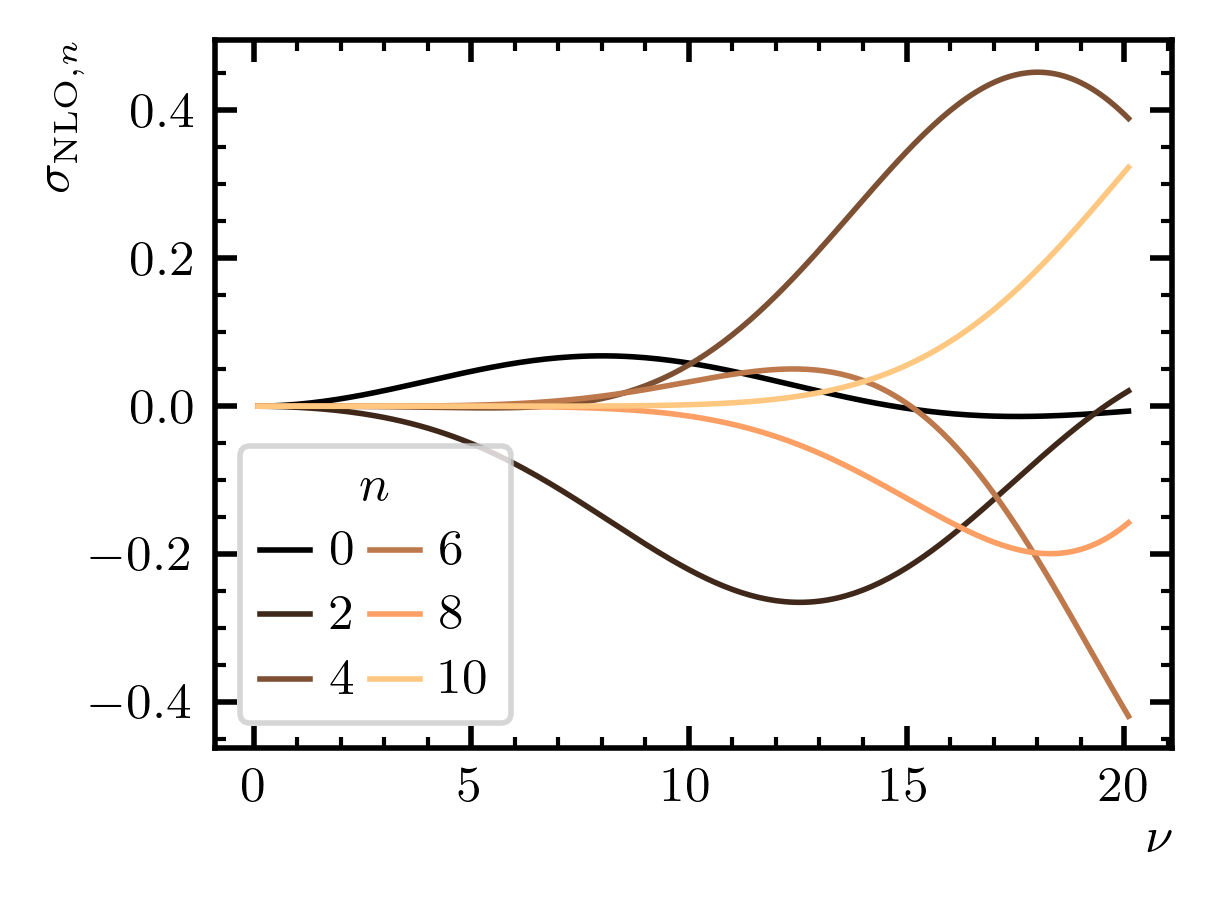}
    \end{subfigure}
    \caption{The \gls{lo} contribution to $\sigma_n$, $\sigma_{\text{LO},n}(\nu)$, and the \gls{nlo} contribution, $\sigma_{\textup{NLO},n}(\nu,z\mu)=\sigma_{n}(\nu,z\mu)-\sigma_{\textup{LO},s}(\nu)$. We used $\lambda=2.7$ and $z_3 = 5 \times \qty{0.0658}{\femto\meter}$ as representative values.}
    \label{fig:sigma_n}
\end{figure}

Now we are in a position to rewrite the relation between the \gls{rpitd} in the continuum and the light-cone \gls{da}, \cref{eq:pseudo-to-light-cone}. In essence, we expand the cosine in a Taylor series, separating the variables $w$, $\nu$ and $x$. Then, we perform the integral over $w$, which yields the moments of the matching kernel $c_{2k}(\nu,z\mu)$, and we expand the powers of $x-1/2$ in terms of Gegenbauer polynomials. This procedure yields
\begin{equation}
    \tilde{\phi}_{\text{lt}}(\nu,z) = \int_0^1 \dd{x} K(x,\nu,z\mu) \phi_{\text{lt}}(x,\mu)
\end{equation}
where the new kernel $K(x,\nu,z\mu)$ is an infinite series
\begin{equation}
    K(x,\nu,z\mu)
    =
    \sum_{n=0}^{\infty} \dfrac{\sigma_{2n}^{(\lambda)}(\nu,z\mu)}{A_{2n}^{(\lambda)}}
    \tilde{G}_{2n}^{(\lambda)}(x).
\end{equation}
The normalization of Gegenbauer polynomials is
\begin{equation}
    A_n^{(\lambda)} = \dfrac{2^{1-2\lambda}\pi\Gamma(n+2\lambda)}{(n+\lambda)\Gamma(\lambda)^2 n!}
\end{equation}
with the gamma functions $\Gamma$. The new set of functions $\sigma_n$ are given by
\begin{equation}
    \label{eq:sigma-n}
    \sigma^{(\lambda)}_n(\nu,z\mu)
    =
    \sum_{k=0}^{\infty} \left(-\dfrac{\nu^2}{4}\right)^k \dfrac{c_{2k}(\nu,z\mu)}{\Gamma(2k+1)} I(n,k,\lambda)
\end{equation}
where $c_{2k}(\nu,z\mu)$ are the moments of the matching kernel and $I(n,k,\lambda)$ is proportional to the Mellin transform of Gegenbauer polynomials,
\begin{equation}
    \label{eq:da-matching-integral}
    I(n,k,\lambda)
    =
    \dfrac{2\pi}{4^{\lambda+k} n!}
    \dfrac{\Gamma(1+2k)\Gamma(n+2\lambda)}{
    \Gamma(\lambda) \Gamma\left(\lambda+\dfrac{n+2k+2}{2}\right) \Gamma\left(1+k-\dfrac{n}{2}\right)
    }.
\end{equation}
The general expression of $I$ simplifies to a beta function for the first values $n=0,2,4,\dots$
Finally, we use the \gls{da} parameterization \cref{eq:light-cone-da-model} such that the integral over $x$ adopts the form of the orthogonality relation between Gegenbauer polynomials. As a consequence, \cref{eq:pseudo-to-light-cone} can be rewritten as
\begin{equation}
    \label{eq:da-leading-twist}
    \tilde{\phi}_{\text{lt}}(\nu,z)
    =
    \sum_{n=0}^{\infty}\tilde{d}_{2n}^{(\lambda)}\sigma_{2n}^{(\lambda)}(\nu,z\mu),
    \qquad\qquad
    \tilde{d}_n^{(\lambda)} = \dfrac{d_n^{(\lambda)}}{4^{\lambda}}.
\end{equation}
In \cref{sec:continuum-limit}, we use \cref{eq:da-leading-twist} to fit the coefficients $d_{2n}$ and the parameter $\lambda$, which determine the \gls{lcda}, to the data obtained from the lattice after taking the continuum limit and removing the higher-twist effects. In the following, it is useful to define the \gls{lo} contribution to $\sigma_n$, which corresponds to the \gls{lo} contribution to the matching kernel,
\begin{equation}
    \sigma_{\textup{LO},n}^{(\lambda)}(\nu)
    =
    \sum_{k=0}^{\infty}\left(-\dfrac{\nu^2}{4}\right)^k \dfrac{I(n,k,\alpha)}{\Gamma(2k+1)}
\end{equation}
as well as the \gls{nlo} contribution,
\begin{equation}
    \sigma_{\textup{NLO},n}^{(\lambda)}
    = \sigma_n^{(\lambda)} - \sigma_{\textup{LO},n}^{(\lambda)}.
\end{equation}
In \cref{fig:sigma_n}, we plot $\sigma_{\textup{LO},n}$ and $\sigma_{\textup{NLO},n}$ as a function of Ioffe time for several values of $n$. Note that all odd $n$'s vanish. Looking at the \gls{lo} contribution, which contributes the most to $\sigma_n$, we observe that each line peaks in a certain domain of Ioffe time and then vanishes. Given that we explore the range $\nu \lesssim 6$ in our calculation, we are most sensitive to $\sigma_0$ and its associated parameter $d_0$ and perhaps $\sigma_2$ and $d_2$, which peaks later. As we shall see, our data can only determine in practice $\lambda$ and $d_0$, and we will set $d_2$, $d_4$, etc.,~to zero. Nonetheless, this will be sufficient to describe our lattice data.

%% file: sections/lattice.tex
\section{Lattice setup}
\label{sec:lattice-setup}

We employ the set of $N_f=2$ \gls{cls} ensembles collected in \cref{tab:cls-ensembles}. These data sets employ the Wilson plaquette gauge action and two mass-degenerate Wilson quarks with non-perturbative $\order{a}$-improvement. The pion masses range between $\qty{190}{\mega\eV} < m_{\Ppi} < \qty{440}{\mega\eV}$, and $\kappa_{\Pcharm}$ is fixed so that $m_{\PDs} = m_{\PDs,\text{phys}}=\qty{1968}{\mega\eV}$ \cite{Workman:2022ynf}. The scale is set using $f_{\PK}$ \cite{Fritzsch:2012wq}, which exhibits a milder mass extrapolation than the pion decay constant.  For more details on the gauge simulations, see \cite{Fritzsch:2012wq,Heitger:2013oaa} and references therein. We employ the package openQCD v2.0 \cite{openqcd20} to compute the quark all-to-all propagators with wall sources diluted in spin. The Dirac equation is solved with deflated \glsxtrshort{sap}-\glsxtrshort{gcr} \cite{Luscher:2003qa,Luscher:2007es,Luscher:2007se}, and the contractions are carried out with a custom version of the \glsxtrshort{ddhmc} algorithm \cite{ddhmc}. Throughout the analysis, statistical errors are propagated using the Gamma method detailed in \cite{Wolff:2003sm,Schaefer:2010hu,Ramos:2018vgu} and implemented in \cite{Joswig:2022qfe}.
\begin{table}
    \centering
    \input{tables/cls-ensembles}
    \caption{The \gls{cls} ensembles used in this study. From left to right the ensemble label, the bare strong coupling, the lattice spacing \cite{Fritzsch:2012wq}, the number of lattice sites in every spatial direction ($T=2L$), the approximate value of the pion mass \cite{DellaMorte:2017dyu}, the proxy of finite-volume effects $m_{\Ppi}L$, the value of $\kappa_{\Pcharm}$ \cite{Balasubramamian:2019wgx} and the number of measurements. Ensembles marked with an asterisk are only used to check the size of \glspl{fse} but are not included in the continuum extrapolation.}
    \label{tab:cls-ensembles}
\end{table}
The momentum $p$ in the final-state meson is introduced using \gls{ptbc} \cite{Sachrajda:2004mi} on one of the charm-quark fields, while the other retains \gls{apbc}. Following the setup described in \cref{sec:pseudo-da-extraction} to isolate $\mathcal{M}$, we apply the twist angle $\theta$ in the z-direction, such that $p_3 L = \theta$ while $p_1=p_2=0$. To obtain a realistic meson $\Petac$ and remove higher excitations, we form a \gls{gevp} with the interpolator $\APcharm\gamma_5\Pcharm$ at four levels of Gaussian smearing \cite{Gusken:1989qx}, which correspond to the smearing radii $r/a=\numlist{0;2.74;3.54;4.47}$. The links in the smearing operator employ ten iterations of \glsxtrshort{ape} smearing \cite{APE:1987ehd} to reduce their short distance fluctuations.
\begin{figure}[htbp]
    \centering
    \begin{subfigure}[t]{0.49\textwidth}
        \centering
        \includegraphics{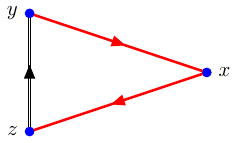}
    \end{subfigure}
    \begin{subfigure}[t]{0.49\textwidth}
        \centering
        \includegraphics{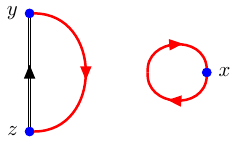}
    \end{subfigure}
    \caption{Wick contractions for the $\Petac$ meson. The double black vector represents a Wilson line and the red solid stripes show the charm-quark propagators. In this work, we only consider the quark-connected contribution on the left.}
    \label{fig:wick-contractions}
\end{figure}
To compute the \gls{da} we need the Wick contractions of $M^\alpha$, which appear in \cref{fig:wick-contractions}. Looking at the disconnected piece first, it induces a mixing with the iso-singlet states $\Peta$ and $\Petaprime$, but in our simulations the latter state cannot appear because the strange-quark is quenched. We exclude this quark-disconnected contribution to simplify the analysis and because we expect it to be small due to \gls{ozi} suppression. In perturbation theory the diagram requires a two-gluon exchange, which means that it is suppressed by a factor $\alpha_s^2(\mu) \sim 0.05$ at our renormalization scale. A single gluon exchange vanishes because this is a vector state, and any expectation value where we project to a pseudoscalar state will vanish, requiring two gluons to obtain the correct quantum numbers.
\begin{figure}[htbp]
    \centering
    \includegraphics{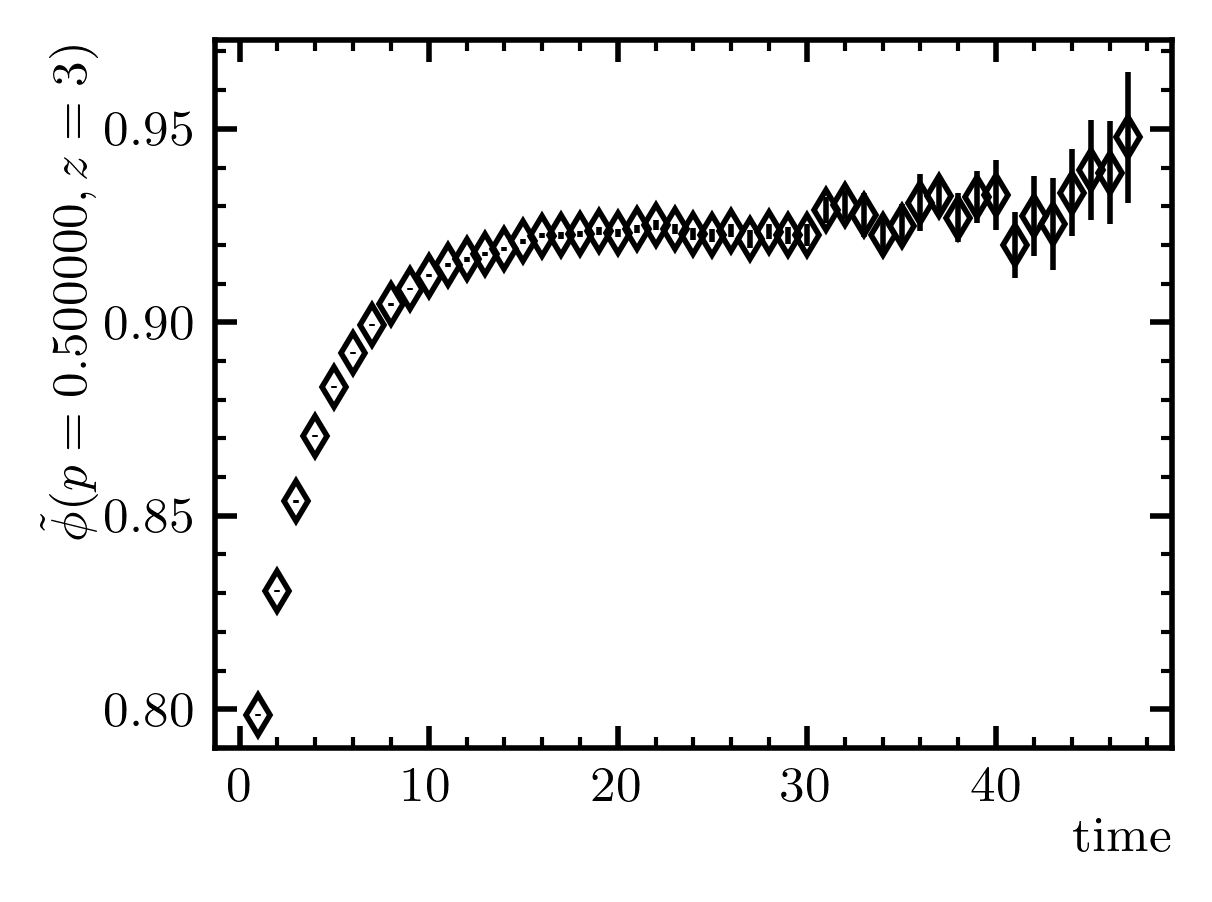}
    \caption{Double ratio in  \cref{eq:rgi-ratio} for parameters $ap=0.5$ and $z_3/a=3$ on ensemble F6. The \gls{rpitd} $\tilde{\phi}(\nu,z)$ can be extracted from the plateau region between time slices 20 and 30.}
    \label{fig:double-ratio}
\end{figure}
\begin{figure}[htbp]
    \centering
    \begin{subfigure}{0.49\textwidth}
        \centering
        \includegraphics{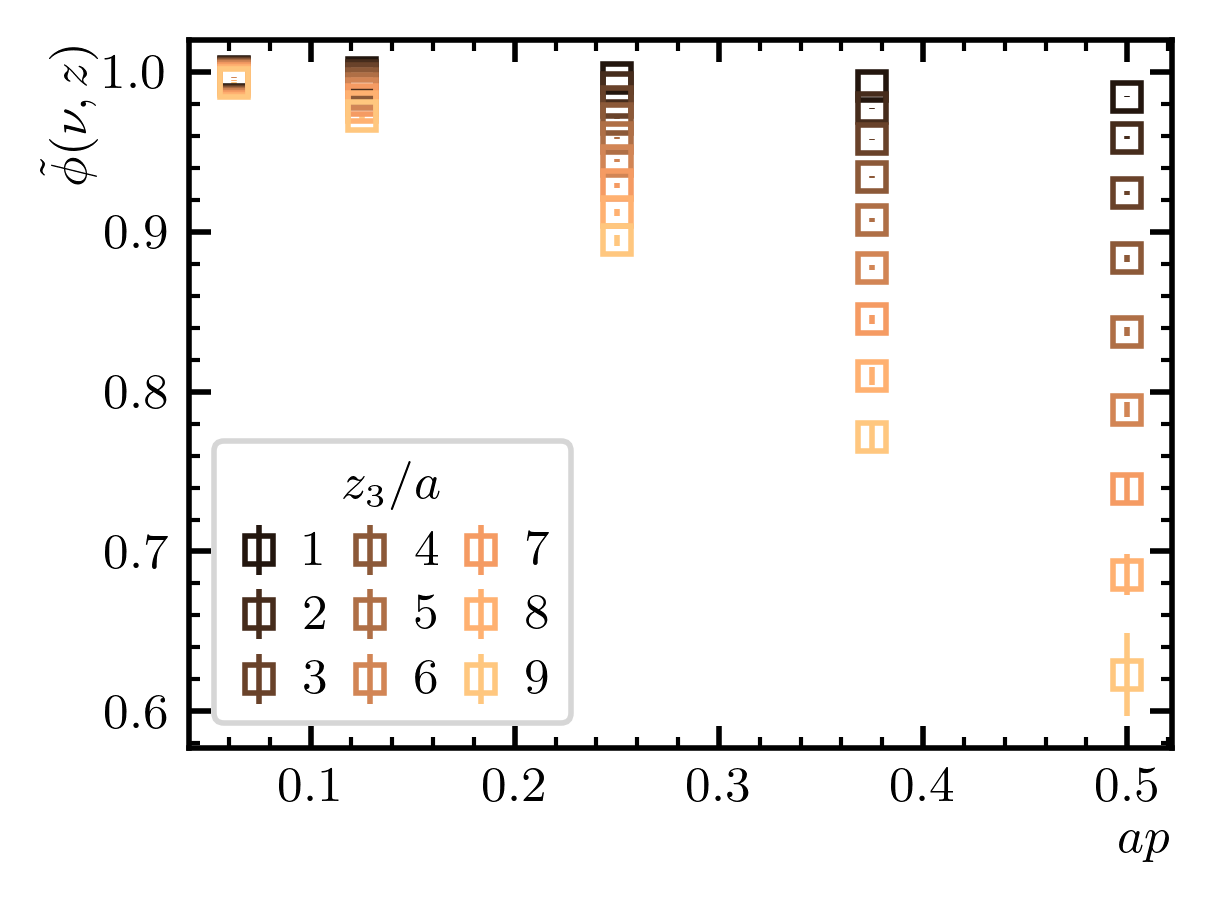}
    \end{subfigure}
    \begin{subfigure}{0.49\textwidth}
        \centering
        \includegraphics{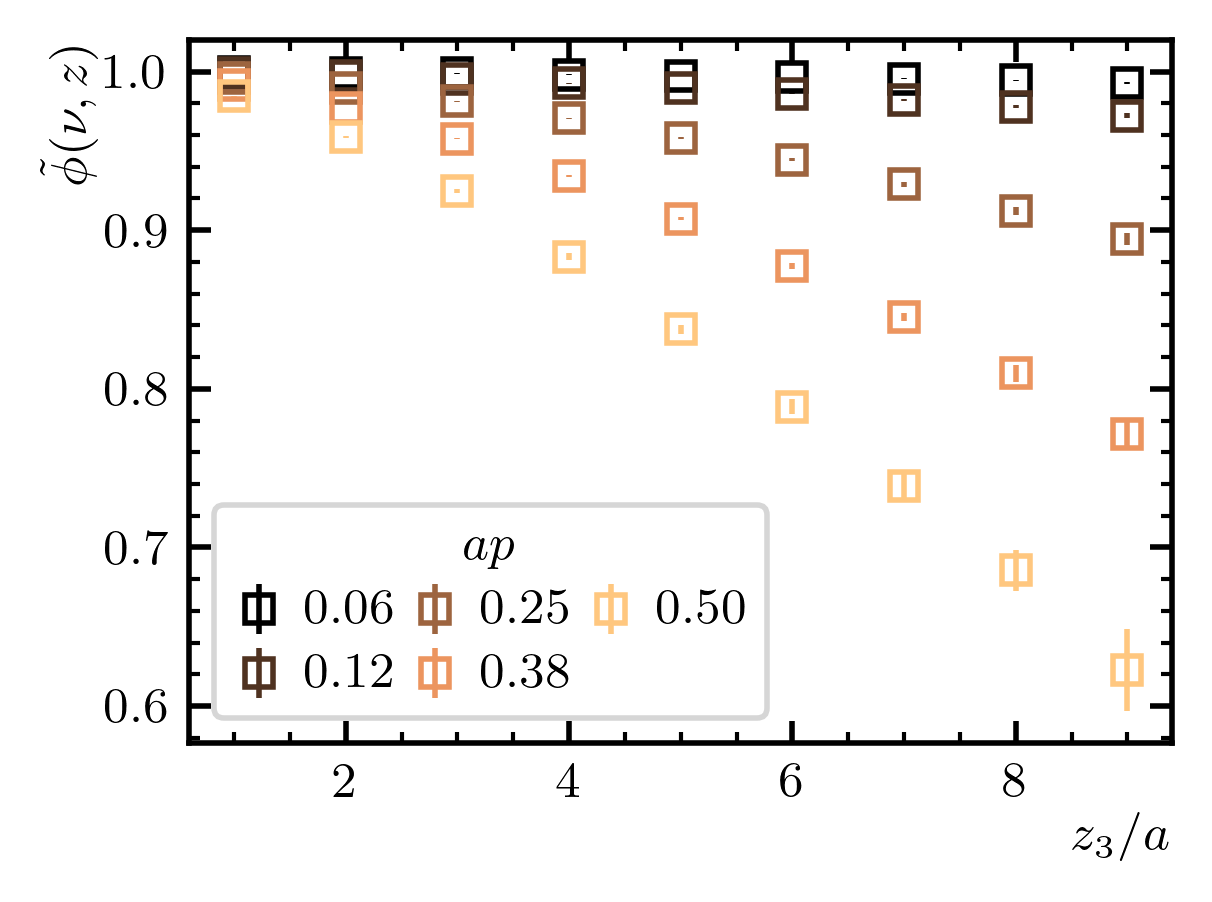}
    \end{subfigure}
    \caption{The \gls{rpitd} defined in \cref{eq:rgi-ratio} computed on ensemble G8 as a function of the final-state meson $p$ and the extension of the Wilson line $z_3/a$.}
    \label{fig:g8-rpitd}
\end{figure}

\begin{figure}[htbp]
    \centering
    \includegraphics[scale=1]{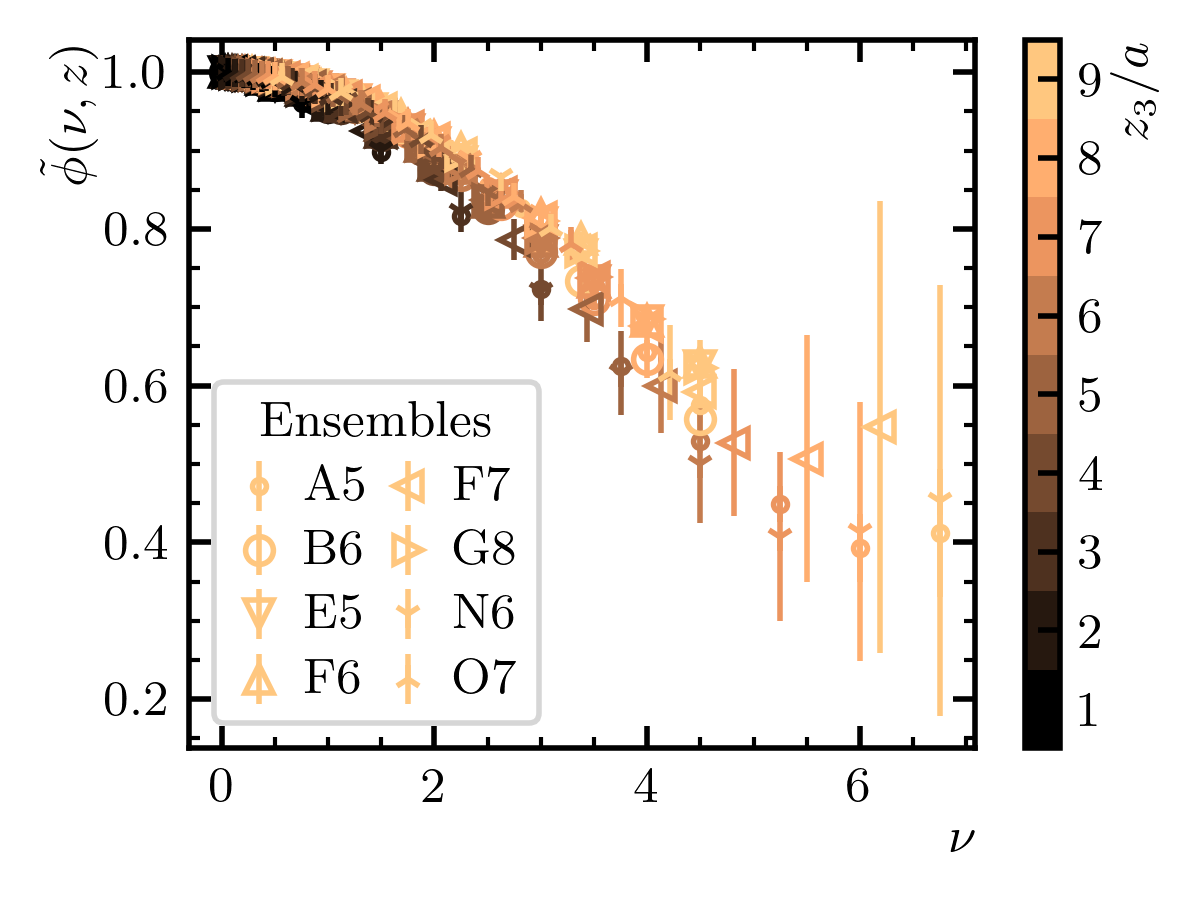}
    \caption{The \gls{rpitd} as a function of Ioffe time on all the ensembles used in the continuum extrapolation. Different ensembles are shown with different markers, and different Wilson lines are given by the color gradient.}
    \label{fig:rpitd}
\end{figure}

The first step of the data analysis is forming \cref{eq:mel-lattice} and the \gls{rgi} ratio \cref{eq:rgi-ratio}. At this stage, the latter shows some time dependence at early times due to excited states, see \cref{fig:double-ratio}. They quickly decay leaving plateaus which are typically $\order{\qty{0.5}{\femto\meter}}$ wide from where we can extract $\tilde{\phi}(\nu,z)$ fitting to a constant. However, one should not forget that all $(\nu,z^2)$ data on a given ensemble are correlated. Tackling the problem head-on, that is, fitting all plateaus in an ensemble simultaneously including their correlations is not possible, as the covariance matrix has a dimension of $\order{1000}$. Our statistics are not sufficient to properly estimate all entries and the matrix is not invertible. Instead, we exploit a hierarchy in the correlation matrix: Data points $(\nu,z^2)$ at the same time slice are far more correlated than points at different time slices. This fact guides our approach to solve the problem. Instead of fitting to a plateau, we select a particular time slice to be $\tilde{\phi}(\nu,z)$ and carry out the entire analysis. This approach preserves the most important correlations in our data while providing a conservative error estimate. Of course, the choice of the time slice introduces a systematic that needs to be assessed (see \cref{sec:systematics}). The estimate of this uncertainty is the second error in \cref{tab:results}. After this step, we obtain the \gls{rpitd} $\tilde{\phi}(\nu,z)$ as given by \cref{eq:rgi-ratio}. The result on ensemble G8 appears in \cref{fig:g8-rpitd} as a function of the final-state momentum $p$ and the extension of the Wilson line. The complete data set, this time as a function of Ioffe time and the Wilson line, appears in \cref{fig:rpitd}. We notice that the data collapses to a nearly universal line, but important lattice artifacts remain which are accounted for in the continuum extrapolation.

%% file: tables/cls-ensembles.tex
\begin{tabular}{lccc cc ccc}
\toprule
  id &
  $\beta$ &
  $a~[\unit{\femto\metre}]$ &
  $L/a$ &
  $am_{\Ppi}$ &
  $m_{\Ppi}~[\unit{\mega\eV}]$ &
  $m_{\Ppi}L$ &
  $\kappa_{\Pcharm}$ &
  Measurements \\
\midrule
  A5  & 5.2 & 0.0755(9)(7) & $32$ & \num{0.1265(8)} & 331 & 4.0 & 0.12531 & 1980 \\
  B6  &     &              & $48$ & \num{0.1073(7)} & 281 & 5.2 & 0.12529 & 1180 \\
\midrule
  D5* & 5.3 & 0.0658(7)(7) & $24$ & \num{0.1499(1)} & 449 & 3.6 & 0.12724 & 1500 \\
  E5  &     &              & $32$ & \num{0.1458(3)} & 437 & 4.7 & 0.12724 & 2000 \\
  F6  &     &              & $48$ & \num{0.1036(3)} & 311 & 5.0 & 0.12713 & 1200 \\
  F7  &     &              & $48$ & \num{0.0885(3)} & 265 & 4.3 & 0.12713 & 2000 \\
  G8  &     &              & $64$ & \num{0.0617(3)} & 185 & 4.1 & 0.12710 & 1790 \\
\midrule
  N6  & 5.5 & 0.0486(4)(5) & $48$ & \num{0.0838(2)} & 340 & 4.0 & 0.13026 & 1900 \\
  O7  &     &              & $64$ & \num{0.0660(1)} & 268 & 4.2 & 0.13022 & 1640 \\
\bottomrule
\end{tabular}

%% file: sections/extrapolation.tex
\section{Continuum limit}
\label{sec:continuum-limit}

\begin{figure}
    \centering
    \includegraphics[scale=1]{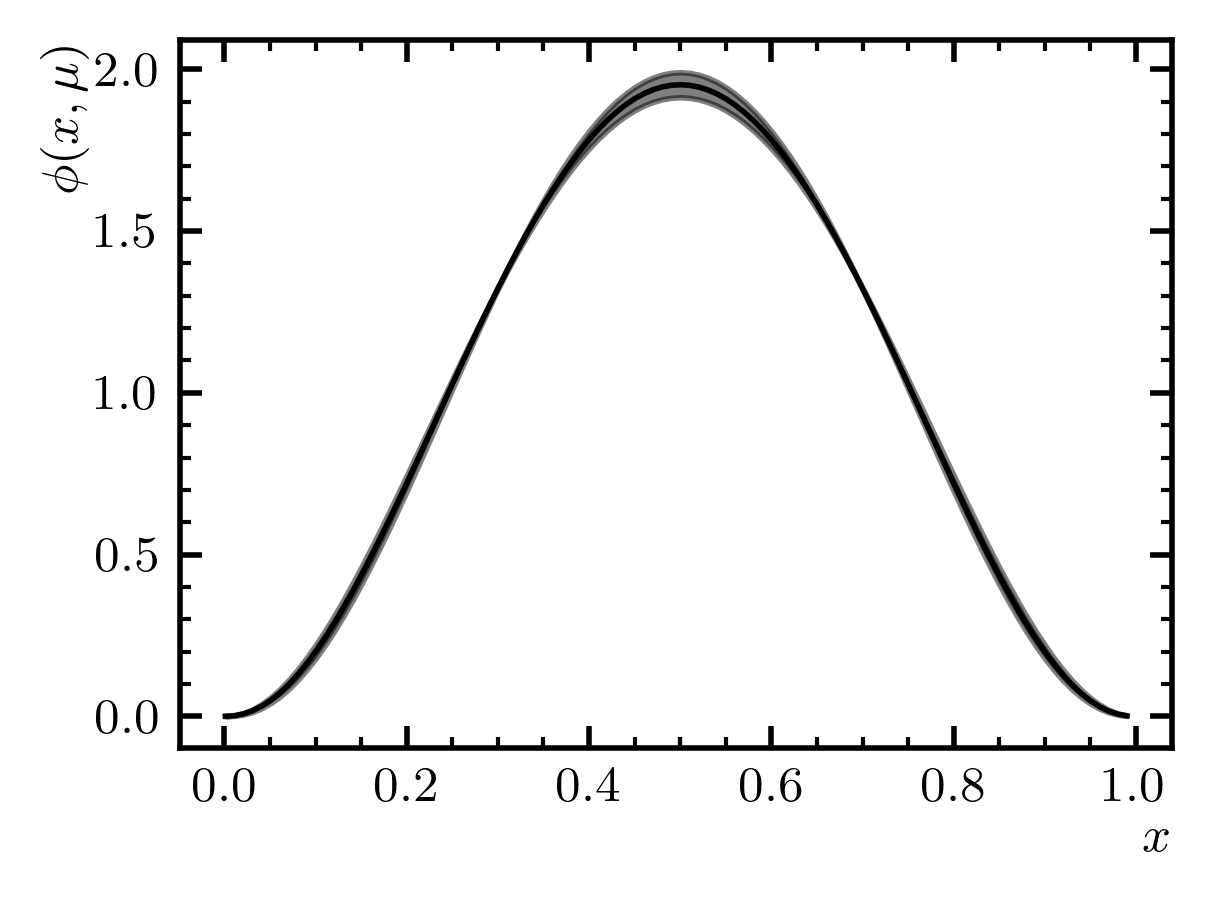}
    \caption{The \gls{lcda} of the $\Petac$-meson. The band shows the statistical error.}
    \label{fig:light-cone-da}
\end{figure}

Once we have obtained $\tilde{\phi}(\nu,z)$ on every ensemble of \cref{tab:cls-ensembles} at various Ioffe times $\nu$ and Wilson lines $z$, we remove the cutoff and match our results to the \gls{lcda} in one single step. See \cite{Karpie:2021pap} for a study of \glspl{pdf} using this approach. Although separating in two distinct steps the extrapolation and the problem with the inverse Fourier transform could be simpler, we would need several lattices for every momentum and Wilson line in physical units. See \cite{Karpie:2019eiq} for alternative methods to recover the \gls{da} from the \gls{itd} with only limited data. The first step to extrapolate to the continuum is building a model to account for the lattice artifacts, the higher-twist contamination, and the small quark mass dependence. We start by noting that the CP symmetry of the strong interactions constraints the behavior of the \gls{da},
\begin{equation}
    \tilde{\phi}(p,z)=\tilde{\phi}^*(-p,z)=\tilde{\phi}^*(p,-z)=\tilde{\phi}(-p,-z).
\end{equation}
In particular, the equality $\tilde{\phi}(p,z)=\tilde{\phi}(-p,-z)$ restricts lattice artifacts with odd powers of $a$ to be functions of $a\abs{p}$ and $a/\abs{z}$. Then, the lattice data $\tilde{\phi}(\nu,z)$ can be related to the continuum \gls{rpitd} $\tilde{\phi}_{\text{con}}(\nu,z)$ via a Taylor expansion in the lattice spacing,
\begin{equation}
  \tilde{\phi}(\nu,z)
  =
  \tilde{\phi}_{\text{con}}(\nu, z)
  +\sum_{r=1}\left(\dfrac{a}{\abs{z}}\right)^rA_r(\nu)
  +\left(a\Lambda\right)^r B_r(\nu),
\end{equation}
where we use $\Lambda\equiv\Lambda^{(2)}_{\text{QCD}}=\qty{330}{\mega\eV}$ \cite{FlavourLatticeAveragingGroupFLAG:2021npn} to render all terms dimensionless. Following \cite{Karpie:2021pap}, we introduce auxiliary functions $A_r$ and $B_r$ to model the Ioffe-time dependence of the lattice artifacts. We define them in analogous way to $\phi(x,\mu)$, and use the same basis of Gegenbauer polynomials,
\begin{equation}
    \begin{aligned}
        A_r^{(\lambda)}(x)
        &=
        (1-x)^{\lambda-1/2} x^{\lambda-1/2}
        \sum_{s=0}^{S_{a,r}} a_{r,2s}^{(\lambda)} \tilde{G}_{2s}^{(\lambda)}(x),
        \\
        B_r^{(\lambda)}(x)
        &=
        (1-x)^{\lambda-1/2} x^{\lambda-1/2}
        \sum_{s=0}^{S_{b,r}} b_{r,2s}^{(\lambda)} \tilde{G}_{2s}^{(\lambda)}(x).
    \end{aligned}
\end{equation}
The unknown functional dependence is contained in the fit parameters $a_{r,2s}$ and $b_{r,2s}$. The function $A_r$ in Ioffe time can be obtained via
\begin{equation}
  A_r^{(\lambda)}(\nu)
  = \int_0^1 \dd{x} A_r^{(\lambda)}(x) \cos\left(x\nu-\nu/2\right)
  = \sum_{s=0}^{S_{A,r}} \tilde{a}_{r,2s}^{(\lambda)} \sigma_{\text{LO},2s}^{(\lambda)}(\nu).
\end{equation}
where $\tilde{a}_{r,2s} = a_{r,2s}/4^{\lambda}$. A similar result holds for $B_r$.
Note that the integrand of $\sigma_{\text{LO},s}$ is even (odd) in the domain of integration for $s$ even (odd), and only the even terms are non-zero. Finally, we set $a_{r,0}=0$ because $\tilde{\phi}(\nu=0,z) = 1$.
The continuum limit $\tilde{\phi}_{\text{con}}$ is a sum of the leading-twist contribution in \cref{eq:da-leading-twist} and higher-twist contamination. We emphasize that the latter piece includes the so-called target-mass corrections, which depend on the physical $\Petac$-meson mass as $z^2 m_{\Petac,\text{phy}}^2$. We model all higher-twist corrections using another auxiliary function, $C_r(\nu)$, which is analogous to $A_r$ and $B_r$,
\begin{equation}
  \tilde{\phi}(\nu,z)
  =
  \tilde{\phi}_{\text{lt}}(\nu,z)
  +\sum_{n=1} \left(\dfrac{a}{\abs{z}}\right)^n A_n(\nu)
  + (a\Lambda)^n B_n(\nu)
  + (z^2\Lambda^2)^n C_n(\nu).
\end{equation}
Note that the target-mass effects are absorbed in the definition of the $C_r(\nu)$ fit parameters. Since we remove them before matching to the light-cone, we need not modify the kernel $C(w,\nu,z\mu)$ to take the meson mass into account.
As the only effect of the light-quark on the \gls{da} comes from the fermionic determinant, we may expect a polynomial expansion in the light-quark mass $m_{\ell}$, which has itself a chiral expansion in $m^2_{\Ppi}$ at \ac{lo}. We checked that the latter function is enough to describe the pion mass dependence for ensembles at $\beta=5.3$. Besides, we take into account the small mistuning in the charm-quark mass with another term proportional to the $\Petac$ mass, $m_{\Petac}$, which we obtain from the \gls{gevp}. After trying several functional combinations, we observe that the type of model which describes best our data is
\begin{equation}
    \begin{aligned}
        \tilde{\phi}(\nu,z)
        &=\tilde{\phi}_{\text{lt}}(\nu,z)
        +\sum_{n=1} 
        \Bigg[
            \left(\dfrac{a}{\abs{z}}\right)^n
            A_n(\nu) + (a\Lambda)^n B_n(\nu) + (z^2\Lambda^2)^n C_n(\nu)
            \\
            &+\left(\dfrac{a}{\abs{z}}\right)^n
            \Big(
            \Lambda^{-1} \left(m_{\Petac}-m_{\Petac,\text{phy}}\right) D_n(\nu)
            + \Lambda^{-2} \left(m_{\Ppi}^2-m_{\Ppi,\text{phy}}^2\right) E_n(\nu)
            \Big)
        \Bigg],
    \end{aligned}
\end{equation}
where we normalize by the physical masses \cite{Workman:2022ynf}
\begin{equation}
    \begin{aligned}
    m_{\Petac,\text{phy}} &= \qty{2983.9(4)}{\mega\eV}, & m_{\Ppi,\text{phy}} &= \qty{134.9768(5)}{\mega\eV}
    \end{aligned}
\end{equation}
and we have introduced further auxiliary functions $D(\nu)$ and $E(\nu)$ with the same form as $A(\nu)$. In practice, our data is only sensitive to the first coefficient $n=1$ in the series of auxiliary functions, the first nonzero coefficient for each function $A(\nu)$, $B(\nu)$, etc., and the first term in $\tilde{\phi}_\text{lt}(\nu,z)$, so that our model simplifies to
\begin{equation}
  \label{eq:continuum-extrapolation-model}
  \begin{aligned}
    \tilde{\phi}(\nu,z)
    &=
    \tilde{\phi}_{\text{lt}}(\nu,z) + \dfrac{a}{\abs{z}} A_1(\nu) + a\Lambda B_1(\nu) + z^2\Lambda^2 C_1(\nu)
    \\
    &+\dfrac{a}{\abs{z}}
      \Big(
        \Lambda^{-1} \left(m_{\Petac} - m_{\Petac,\text{phy}}\right) D_1(\nu)
        + \Lambda^{-2} \left(m_{\Ppi}^2 - m_{\Ppi,\text{phy}}^2\right) E_1(\nu)
      \Big).
  \end{aligned}
\end{equation}
To fit \cref{eq:continuum-extrapolation-model} to our lattice data we minimize a chi-square using the \gls{varpro} algorithm ---see \cite{doi:10.1137/0710036} for the original work, and \cref{sec:varpro} for our particular implementation. We include the correlations among the data on each ensemble, and tame very small eigenvalues in the covariance matrix using the averaging method outlined in \cite{Michael:1994sz} and implemented in \cite{Joswig:2022qfe}. We find the minimum at $\chi^2/\text{dof}=368/467=0.79$ and the results are gathered in \cref{tab:results}. Looking at its second column, where the fit parameter estimates appear, we observe that $\lambda$ lies far away from its asymptotic value of 1.5. Furthermore, we observe a certain hierarchy in the coefficients: Those associated with terms proportional to $a/\abs{z}$ and the mistuning of the charm-quark mass are the most relevant, followed by pure lattice artifacts, the pion mass dependence and higher-twist effects. Nonetheless, every term is necessary to describe the data well, and in particular we observe non-zero higher-twist effects, which include target-mass corrections. See \cref{sec:systematics} for an explanation of the two systematic errors.
\begin{table}
    \centering
    \input{tables/results}
    \caption{Expected value and uncertainty for each fit parameter. The first column labels the parameter. The second column corresponds to a global fit to all \cref{tab:cls-ensembles} ensembles except D5. The first uncertainty in this column indicates the statistical error, the second shows the variation of our result depending on the time slice selected as $\tilde{\phi}(\nu,z)$ (see \cref{sec:lattice-setup}), and the third the extrapolation uncertainty. The third column removes the heavier pion mass, E5, from the fit, and the fourth column includes only ensembles with $m_{\Ppi}<\qty{300}{\mega\eV}$. In the fifth column, we remove all Wilson lines with $z_3/a > \qty{0.5}{\femto\meter}$. For details about the systematic errors and this table, see \cref{sec:systematics}.}
    \label{tab:results}
\end{table}
Now we can evaluate the \gls{lcda} defined in \cref{eq:light-cone-da-model} using the first term in the series and $\lambda=\num{2.73(18)}$,
\begin{equation}
    \label{eq:main-result}
    \phi_{\text{lt}}(x,\mu) = \dfrac{4^{\lambda} (1-x)^{\lambda-1/2} x^{\lambda-1/2}}{B\left(1/2, 1/2+\lambda\right)},
\end{equation}
where we used the fact that $\tilde{G}_{0}(x)=1$. The plot of $\phi(x,\mu)$ can be seen in \cref{fig:light-cone-da}. We tried to add higher coefficients to describe the \gls{da}, in particular $d_2$ and $d_4$, which correspond to $\tilde{G}_{2}(x)$ and $\tilde{G}_{4}$, respectively. Although they are both compatible with zero, our data has not sufficient range in $\nu$ to determine them reliably.

%% file: tables/results.tex
\begin{tabular}{crrrr}
    \toprule
    $\lambda$
    & $\num[parse-numbers=false]{2.73\pm0.12\pm0.12\pm0.06}$
    & $\num{2.75\pm0.12}$
    & $\num{2.61\pm0.15}$
    & $\num{2.62\pm0.10}$
    \\
    $a_{1,2}$
    & $\num[parse-numbers=false]{-7.58\pm0.05\pm0.59\pm0.55}$
    & $\num{-8.12\pm0.05}$
    & $\num{-8.68\pm0.13}$
    & $\num{-6.76\pm0.04}$
    \\
    $b_{1,2}$
    & $\num[parse-numbers=false]{0.88\pm0.07\pm0.08\pm0.06}$
    & $\num{0.89\pm0.07}$
    & $\num{0.77\pm0.10}$
    & $\num{0.81\pm0.07}$
    \\
    $c_{1,2}$
    & $\num[parse-numbers=false]{-0.042\pm0.002\pm0.005\pm0.001}$
    & $\num{-0.0428\pm0.0022}$
    & $\num{-0.0440\pm0.0028}$
    & $\num{-0.0407\pm0.0024}$
    \\
    $d_{1,2}$
    & $\num[parse-numbers=false]{-2.221\pm0.015\pm0.063\pm0.15}$
    & $\num{-2.368\pm0.015}$
    & $\num{-2.52\pm0.04}$
    & $\num{-2.000\pm0.011}$
    \\
    $e_{1,2}$
    & $\num[parse-numbers=false]{-0.0897\pm0.001\pm0.159\pm0.116}$
    & $\num{-0.1700\pm0.0016}$
    & $\num{-0.321\pm0.005}$
    & $\num{-0.06848\pm0.00012}$
    \\
    \bottomrule
\end{tabular}

%% file: sections/systematics.tex
\section{Systematics}
\label{sec:systematics}

Lattice simulations are performed in a finite volume, but it is only after estimating the infinite volume limit that we can compare our results to physical quantities. The associated difference between the two results, or \glspl{fse}, are especially difficult to estimate in the case of concern here, a non-local matrix element. Some works have explored this problem, and for example the authors of \cite{Briceno:2018lfj} give analytic expressions for a non-local matrix element with scalar currents. Although it omits the Wilson line, one can see that the main \glspl{fse} stem from the pion as a decreasing exponential of $m_{\Ppi}L$, plus some extra function for the $\Petac$ in terms of $m_{\Petac}(L-z)$. Their subsequent work in \cite{Briceno:2021jlb} determines the same quantity non-perturbatively in terms of the form factors of the asymptotic hadron state. Yet, the estimation of \glspl{fse} for non-local matrix elements remains a complicated subject, and no theoretical prediction exists for the \gls{da}. Several studies of the nucleon \glspl{pdf} report negligible effects \cite{PhysRevD.100.074502,Alexandrou:2019lfo}, while other detect them with some significance \cite{Joo:2019jct}. In the case of the pion \gls{pdf}, the authors of \cite{Joo:2019bzr} perform a fit to capture the volume effects directly. In our calculation, we compare the results for the \gls{rpitd} $\tilde{\phi}(\nu,z)$ on ensembles D5 and E5, which only differ by their volume, and found very good agreement. Therefore, we neglect \glspl{fse} in our project and do not attempt to correct our data.

As already mentioned in \cref{sec:lattice-setup}, an important source of systematic uncertainty originates in the extraction of $\tilde{\phi}(\nu,z)$ from the lattice data. The ratio in \cref{eq:rgi-ratio} shows a residual time dependence, although a wide plateau is visible for all the momenta and Wilson lines. For a given ensemble, we choose one time slice as the \gls{rpitd} $\tilde{\phi}(\nu,z)$ (the same for all $(\nu,z^2)$) and extrapolate to the continuum. In this way we can keep the correlations between the different Ioffe times and Wilson lines intact, all while the correlation matrix remains invertible. Accounting for all the possible time slices that can be chosen on the various ensembles, there are more than 3.5 million possible extrapolations to the continuum. To estimate the impact of our choice on the final result, we sample nearly 16000 cases. The spread of the results, which is Gaussian, gives the second uncertainty in \cref{tab:results}.

To further test the stability of the continuum extrapolation we perform two cuts in the pion mass. First, we fit only ensembles with $m_{\Ppi} < \qty{400}{\mega\eV}$, which removes ensemble E5. Second, we perform a more stringent cut and we only fit $m_{\Ppi} < \qty{300}{\mega\eV}$, which further removes A5, F6 and N6. The result of both fits appear in \cref{tab:results} together with the fit that considers all ensembles. We observe that the physical parameter $\lambda$ remains compatible while some of the nuisance parameters have a more pronounced shift, especially $e_{1,2}$ that takes care of the pion-mass dependence. We employ the mass cuts to estimate the systematic uncertainty associated to the extrapolation, which is fixed to half the distance between the fit to all ensembles and the cut furthest away. This is the third uncertainty in \cref{tab:results}.

The perturbative matching we use in \cref{sec:pseudo-da-extraction} is only valid for sufficiently short Wilson lines. In particular, \cite{Radyushkin:2019mye} argues that $z_3 \leq \qty{0.5}{\femto\meter}$ is necessary. In our analysis, see for instance \cref{fig:rpitd}, five Wilson lines exceed this mark: $z_3/a > 6$ for $\beta = 5.2$ and $z_3/a > 7$ for $\beta = 5.3$. To test the validity of our results, we repeat the extrapolation to the continuum as given in \cref{sec:continuum-limit}, which also includes the matching procedure, removing these Wilson lines. The result we obtain with this reduced data set is $\lambda = \num{2.62(10)}$, $\chi^2/\text{dof} = 388/374=1.04$. The uncertainty is slightly smaller since the data with the largest $z^2$ are noisier than average. We report the value of all parameters in this fit on the fifth column of \cref{tab:results}. The value of $\lambda$ is compatible with our preferred result (see second column of \cref{tab:results}), and we use this as a corroboration of our analysis. We decide against adding the difference between the two results as an extra systematic uncertainty, and we rather see the proximity of both as proof that, in our case, it is safe to include slightly longer Wilson lines. Regarding the nuisance parameters, the situation is similar, with the variation well explained by our predetermined error budget. In particular, the value of $c_{1,2}$, which controls the size of higher-twist corrections, is unchanged within errors.

Before finishing this section, we note several other systematic uncertainties that we cannot estimate at the moment. Most important of all, the strange quark is quenched, and its effects on our determination can only be estimated reliably including it in the action. Second, an extra lattice spacing would allow us to estimate the $\order*{a^2}$ dependence neglected in our continuum extrapolation. Finally, we consider the pion mass dependence in our results, and although it is only a subleading effect, having an ensemble at the physical pion mass would remove any systematic associated with the extrapolation in the mass.

%% file: sections/comparisons.tex
\section{Comparison to other approaches}
\label{sec:comparisons}

As a last step, we compare our \textit{ab initio} determination of the \gls{da} to the \gls{nrqcd} calculation in \cite{Chung:2019ota} and the \gls{ds} method used in \cite{Ding:2015rkn}. The former alternative assumes that the charm-quarks can be approximated as non-relativistic particles, whose \gls{da} at \gls{lo} is a Dirac delta centered at $x=1/2$. The authors of \cite{Chung:2019ota} further include the first relativistic and quantum corrections, which they note are of the same size, to produce a more accurate description. In \cref{sec:nrqcd}, we gather the important equations for this method. The latter alternative \cite{Ding:2015rkn} solves the \gls{ds} equations to obtain Bethe-Salpeter wave functions \cite{Chang:2013pq,Gao:2014bca}, which are related to the \gls{da}. The data on the light cone are subsequently fitted to a finite-width representation of a Dirac delta, which appears together with the other important parameters of this method in \cref{sec:ds}. To compare to the latter, we also need to evolve the \gls{da} in \cite{Ding:2015rkn} from their original scale, $\tilde{\mu}=\qty{2}{\giga\eV}$, to ours, $\mu=\qty{3}{\giga\eV}$. To solve the evolution equations we employ the APFEL++ suite \cite{Bertone:2013vaa,Bertone:2017gds}.

\begin{figure}
    \centering
    \includegraphics{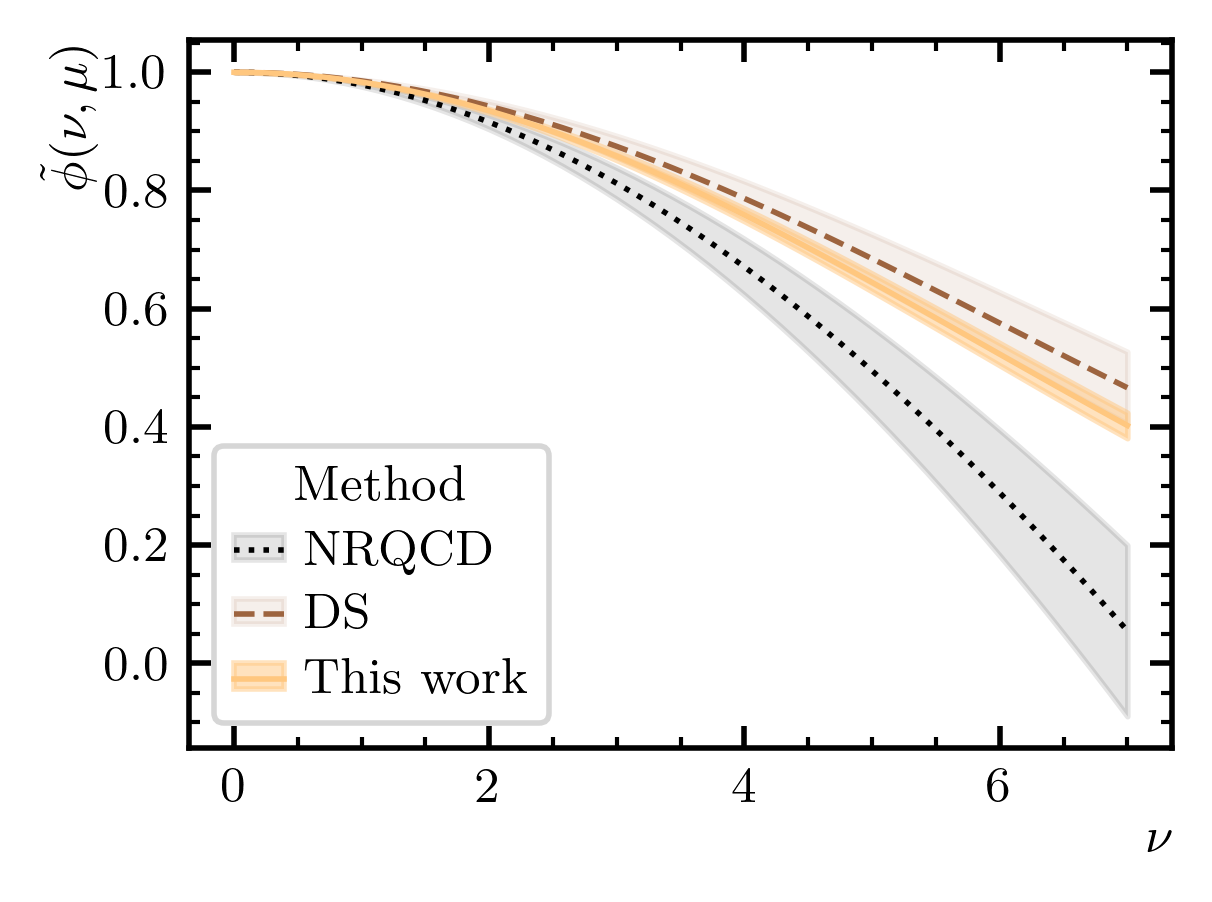}
    \caption{Comparison in Ioffe-time space between the \gls{da} as determined in this work, by \gls{nrqcd} \cite{Chung:2019ota} and by \gls{ds} \cite{Ding:2015rkn}. The bands reflect the uncertainties reported on each work, ours indicating the total error.}
    \label{fig:comparisons}
\end{figure}

Especially for \gls{nrqcd}, it is interesting to Fourier-transform to Ioffe-time space and compare the various determinations, getting rid of the Dirac delta and its derivatives appearing in $x$ space. The Fourier transform of the \gls{nrqcd} and \gls{ds} \glspl{da} appears in \cref{sec:nrqcd,sec:ds}, while the Fourier transform of our \gls{da} is simply given by
\begin{equation}
    \tilde{\phi}_{\text{lt}}(\nu,\mu) = \sum_{n=0}^{\infty} \tilde{d}_{2n}\sigma_{\textrm{LO},2n}(\nu).
\end{equation}
The three \glspl{da} appear in \cref{fig:comparisons}, and we observe excellent agreement between the lattice and the \gls{ds} determination, while both show a very different behavior from \gls{nrqcd} for larger Ioffe times. The band estimates the uncertainty of each determination: For our result, it is the total error; for \gls{nrqcd}, it is the uncertainty of $\expval*{v^2}$, a parameter defined in \cref{sec:nrqcd}; and for \gls{ds}, it indicates the deviation between the data points obtained from the \gls{ds} equations and the functional form used to fit them. Furthermore, observe that at leading order the \gls{nrqcd} \gls{da} is equal to 1, and the first-order corrections are just as large around $\nu \sim 6$, with the quantum loops and the relativistic correction of similar size. Therefore, we argue that one should extend the \gls{nrqcd} calculation to the next order in both quantum and relativistic corrections.

Finally, we compute the first Mellin moments for each determination,
\begin{equation}
    \expval*{g^n}(\mu) = \int_0^1 \dd{x}g^n\phi(x,\mu),
\end{equation}
where $g = -1+2x$ and only even $n$'s are non-zero. \Cref{sec:da-moments,sec:nrqcd,sec:ds} gather the expressions for the moments, and we show $\expval*{g^2}$ and $\expval*{g^4}$ in \cref{tab:moments}. Once more, we find good agreement only with the \gls{ds} method. Looking at the Mellin moments of \gls{nrqcd} in \cref{sec:nrqcd}, we see that $\expval*{v^2}$ only contributes for $\expval*{g^2}$, and further corrections should be included to estimate $\expval*{g^4}$. This explains the tiny uncertainty in $\expval*{g^4}$.
\begin{table}
    \centering
    \input{tables/moments}
    \caption{First nonzero moments of the \gls{da} as determined by this work, by the \gls{ds} approach in \cite{Ding:2015rkn}, and by the \gls{nrqcd} calculation in \cite{Chung:2019ota}. We define $g=-1+2x$ for $x\in[0,1]$.}
    \label{tab:moments}
\end{table}

%% file: tables/moments.tex
\begin{tabular}{c ccc}
\toprule
                & This work      & \gls{ds}        & \gls{nrqcd}       \\
$\expval*{g^2}$ & \num{0.134(6)} & \num{0.118(18)} & \num{0.171(23)}    \\
$\expval*{g^4}$ & \num{0.043(4)} & \num{0.036(9)}  & \num{0.018808(19)} \\
\bottomrule
\end{tabular}

%% file: sections/conclusions.tex
\section{Conclusions}
\label{sec:conclusions}

We present the first lattice calculation of the $\Petac$-meson \gls{da}, which is parametrized in \cref{eq:main-result}. We compute the \glsentrylong{pda} on the lattice and match it to the light-cone \gls{da}, which we study via a series of Gegenbauer polynomials that allows for systematic improvement upon extending the data to larger Ioffe times. We employ three different lattice spacings and a wide range of pion masses, allowing us to take the continuum limit accounting for the subleading effects of the quark masses. We study several systematics, and in particular we find that \glspl{fse} are negligible. We compare our results in the continuum to other approaches, and find good agreement with the \glsxtrlong{ds} equations, while we strongly disagree with the \gls{nrqcd} prediction at all but the smallest Ioffe times. In the future we will address several sources of systematics with a new set of simulations at the physical pion mass which include also the strange and charm quarks in the sea.

%% file: appendices/varpro.tex
\section{Variable projection}
\label{sec:varpro}

All but one fit parameters in \cref{eq:continuum-extrapolation-model} are linear. This suggests that we should separate the extrapolation to the continuum limit in a nonlinear minimization, where only $\lambda$ is optimized, and a linear minimization, where all the linear fit parameters are obtained from the optimal value of $\lambda$. This idea is commonly known as \glsfirst{varpro}, it was originally suggested in \cite{doi:10.1137/0710036}, and implementations for different types of problems exist in Python \cite{math11132839}, Matlab \cite{oleary:2013eg} and other languages. See \cite{golub2003separable} for a review of the general concept and applications. In this section, we present our implementation. Start from the common chi-square
\begin{equation}
    \chi_1^2 = \left(\bar{y}-\eta\right)^{\tmat} C^{-1} \left(\bar{y}-\eta\right)
\end{equation}
where the column vector $\bar{y}\in\mathbb{R}^{n \times 1}$ contains the data, $\eta\in\mathbb{R}^{n\times 1}$ is the model, and $C\in\mathbb{R}^{n \times n}$ is the covariance matrix. For example, if we were fitting ensembles B6, F7 and O7, the data vector would be
\begin{equation}
  \bar{y}
  =
  \begin{pmatrix}
    \tilde{\phi}_{\text{B6},0}, & \dots & \tilde{\phi}_{\text{B6},N_{\text{B6}}-1}, &
    \tilde{\phi}_{\text{F7},0}, & \dots & \tilde{\phi}_{\text{F7},N_{\text{F7}}-1}, &
    \tilde{\phi}_{\text{O7},0}, & \dots & \tilde{\phi}_{\text{O7},N_{\text{O7}}-1}
  \end{pmatrix}^{\tmat},
\end{equation}
where the dots run over all datapoints $\tilde{\phi}$ for a particular ensemble and $n=N_{\text{B6}}+N_{\text{F7}}+N_{\text{O7}}$. In the same fashion, the covariance matrix would be block diagonal,
\begin{equation}
  C = C_{\text{B6}} \oplus C_{\text{F7}} \oplus C_{\text{O7}},
\end{equation}
and the chi-square $\chi_1^2$ fits each data point $\bar{y}_i$ for $i=0,\dots,n-1$ using the model
\begin{equation}
  \eta_i
  =
  \sum_{j=0}^{q-1} \mathfrak{q}_{j} \Phi_{i,j}(\mathfrak{r}) + \Theta_i(\mathfrak{r})
\end{equation}
which has $q$ linear fit parameters $\mathfrak{q}\in\mathbb{R}^{q \times 1}$ and $r$ nonlinear fit parameters $\mathfrak{r}\in\mathcal{S_{\mathfrak{r}}}\subset\mathbb{R}^{r \times 1}$, all organized in column vectors. $\mathcal{S_{\mathfrak{r}}}$ is the subspace of values that the parameters $\mathfrak{r}$ are allowed to adopt given the physical constraints of the problem. $\Phi(\mathfrak{r})\in\mathbb{R}^{n \times q}$ is a matrix of nonlinear functions of $\mathfrak{r}$, which is a mapping between the latter and a linear transformation $\mathcal{L}(\mathbb{R}^n,\mathbb{R}^q)$. $\Theta_i(\mathfrak{r})$ is a column vector of similar properties to $\Phi(\mathfrak{r})$. Using the extrapolation model \cref{eq:continuum-extrapolation-model} as an example, $\mathfrak{r}$ has only one entry,
\begin{equation}
    \mathfrak{r}
    =
    \begin{pmatrix}
    \lambda
    \end{pmatrix}.
\end{equation}
If we expand $\tilde{\phi}_{\text{lt}}$ up to $n=4$ and we include several nuisance functions with only one parameter, the linear fit parameters are
\begin{equation}
  \mathfrak{q}
  =
  \begin{pmatrix}
    d_2, & d_4, & a_{1,2}, & b_{1,2}, & c_{1,2}
  \end{pmatrix}^{\tmat}
\end{equation}
and the columns of $\Phi$ and $\Theta$ contain the information about
\begin{equation}
    \begin{aligned}
        \Phi_i
        &=
        \begin{pmatrix}
            \sigma_2, & \sigma_4,
            & \dfrac{a}{\abs{z}}\sigma_{\text{LO},2},
            & a\Lambda\sigma_{\text{LO},2},
            & z^2\Lambda^2\sigma_{\text{LO},2}
        \end{pmatrix},
        &
        \Theta_i
        &=
        \begin{pmatrix}
            \dfrac{4^{\lambda}\sigma_0}{B(1/2,\lambda+1/2)}
        \end{pmatrix}
    \end{aligned}
\end{equation}
where each row evaluates a given point $(\nu,z)$. We suppose the minimum chi-square is obtained for certain values $(\hat{\mathfrak{r}},\hat{\mathfrak{q}})$,
\begin{equation}
  \min_{\mathfrak{r},\mathfrak{q}}\chi_1^2 = \chi_1^2(\hat{\mathfrak{r}},\hat{\mathfrak{q}}).
\end{equation}
When building the covariance matrix $C$, we control its small eigenvalues using the averaging procedure given in \cite{Michael:1994sz}. Then, we can decompose the covariance matrix, $C^{-1} = (L^{-1})^{\tmat} L^{-1}$, define new variables
\begin{align}
  A &\equiv L^{-1} \Phi \in \mathbb{R}^{n \times q},
  &
  B &\equiv L^{-1} \Theta \in \mathbb{R}^{n \times 1},
  &
  y &\equiv L^{-1} \left(\bar{y}-\Theta\right) \in \mathbb{R}^{n \times 1},
\end{align}
and rewrite the chi-square problem as a 2-norm,
\begin{equation}
  \label{eq:chi-square-no-errors}
  \chi_1^2 = \norm{y-A\mathfrak{q}}_2^2.
\end{equation}
Rewriting our problem in the form of \cref{eq:chi-square-no-errors} enables us to use the results of \cite{doi:10.1137/0710036}, although we must take into account that $y$ is also a function of the nonlinear parameters $\mathfrak{r}$. For a given value of $\mathfrak{r}$, the minimum of $\chi_1^2$ occurs at
\begin{equation}
  \hat{\mathfrak{q}} = A^{+}y,
\end{equation}
where the generalized left inverse of $A$ is given by
\begin{equation}
  A^{+} = \left(A^{\tmat}A\right)^{-1}A^{\tmat}.
\end{equation}
The form of $A^{+}$ is problematic because we need to compute the inverse of a large $n \times n$ matrix which can be singular or almost singular. Fortunately, we may choose other definitions for $A^{+}$ because the property it needs to fulfill is $AA^{+}A=A$. To improve stability, we choose the \gls{svd}
\begin{equation}
  A = USV^{\dagger} \to A^{+} = VS^{-1}U^{\dagger},
\end{equation}
where $U$ and $V$ are unitary matrices, and $S$ is a diagonal matrix containing the singular values of $A$.
In our case, $A$ is real and has not full rank, and so $U\in\mathbb{R}^{n \times q}$, $S\in\mathbb{R}^{q \times q}$, and $V\in\mathbb{R}^{q \times q}$.
Just as in \cite{doi:10.1137/0710036}, we can obtain the optimal parameters $\hat{\mathfrak{r}}, \hat{\mathfrak{q}}$ with the following algorithm:
\begin{enumerate}
  \item Obtain the optimal values $\hat{\mathfrak{r}}$ minimizing the modified norm
  \begin{equation}
    \begin{aligned}
      \chi_2^2(\mathfrak{r}) &= \norm{P^{\perp}y}_2^2,
      &
      P^{\perp} = 1 - AA^{+}.
    \end{aligned}
  \end{equation}
  \item Obtain the optimal fit parameters $\hat{\mathfrak{q}}$ using
  \begin{equation}
    \hat{\mathfrak{q}} = A^{+}(\hat{\mathfrak{r}})y(\hat{\mathfrak{r}}).
  \end{equation}
  \item The fit quality is given by
  \begin{equation}
    \min_{\mathfrak{r},\mathfrak{q}}\chi_1^2
    =
    \chi_1^2(\hat{\mathfrak{r}},\hat{\mathfrak{q}})
    =
    \min_{\mathfrak{r}}\chi_2^2
    =
    \chi_2^2(\hat{\mathfrak{r}}).
  \end{equation}
\end{enumerate}
For the minimization of $\chi_2^2$ we need its gradient with respect to $\mathfrak{r}$, which is given by \cite{doi:10.1137/0710036}
\begin{equation}
  \nabla \chi_2^2(\lambda) = y^{\tmat}\nabla P^{\perp} y + 2y^{\tmat} P^{\perp} \nabla y.
\end{equation}
The last term does not appear in \cite{doi:10.1137/0710036} because they do not consider the possibility of an affine fit function $\Theta$. The derivative of the projector is
\begin{equation}
  \nabla P^{\perp}
  =
  - \left(P^{\perp}\nabla A A^{+}+\left(P^{\perp}\nabla AA^{+}\right)^{\tmat}\right)
\end{equation}
while the derivative of the shifted datapoints is simply
\begin{equation}
  \nabla y = -L^{-1}\nabla\Theta.
\end{equation}

%% file: appendices/moments.tex
\section{Moments of the DA}
\label{sec:da-moments}

The Mellin moments of the parameterization in \cref{eq:light-cone-da-model} are
\begin{equation}
    \expval{g^n} = 
    \begin{cases}
        0 & \text{if n is odd},\\
        \dfrac{1}{4^{\lambda}} \sum_{j=0}^{\lfloor n/2 \rfloor} d_{2j}^{(\lambda)}
        I\left(2j, \frac{n}{2}, \lambda\right)
        & \text{if n is zero or even},
    \end{cases}
\end{equation}
where $I$ is defined in \cref{eq:da-matching-integral}, and $n=0,1,2,\dots$, $2j=0,1,2,\dots,n$, and $\lambda>-1/2$ but $\lambda \neq 0$. Using our result $\lambda=\num{2.73(0.18)}$ and neglecting the parameters $d_2=d_4=\dots=0$, the first couple of nonzero \gls{da} moments are
\begin{equation}
    \begin{aligned}
        \expval{g^2} &= \dfrac{I(1,0,\lambda)}{B\left(\dfrac{1}{2}, \dfrac{1}{2}+\lambda\right)} = \num{0.134+-0.006}, &
        \expval{g^4} &= \dfrac{I(2,0,\lambda)}{B\left(\dfrac{1}{2}, \dfrac{1}{2}+\lambda\right)} = \num{0.043+-0.004}.
    \end{aligned}
\end{equation}
We are making a clear systematic error in the determination of the moments, since we are not able to measure $d_2$, $d_4$, etc. However, the method is systematically improvable as we extend the domain of data to larger Ioffe times.

The moments $\expval*{g^n}$ and the Gegenbauer moments $d_n^{(3/2)}$ are related by
\begin{equation}
    \expval{g^n} = A_n+\sum_{j=1}^n d_n^{(3/2)} B_{n,j}
\end{equation}
where
\begin{equation}
    A_n=
    \begin{cases}
        0 & \text{if n is odd}, \\
        \dfrac{3}{2} \left(\dfrac{1}{1+n}-\dfrac{1}{3+n}\right) & \text{if n is even or zero},
    \end{cases}
\end{equation}
and
\begin{equation}
    B_{n,j}=
    \begin{cases}
        \dfrac{3}{2^{n+3}} \dfrac{\sqrt{\pi}(j+2)(j+1)n!}{\Gamma\left(\dfrac{5}{2}+\dfrac{n+j}{2}\right)\Gamma\left(1+\dfrac{n-j}{2}\right)} & \text{if}~n+j~\text{is even and}~j \leq n, \\
        0 & \text{otherwise}
    \end{cases}
\end{equation}
The first few nonzero values of the coefficients $A_n$ and $B_{n,j}$ appear in \cref{tab:da-gegenbauer-moments-coefficients}, and the first nonzero Gegenbauer moments are
\begin{equation}
    \begin{aligned}
        d_2^{(3/2)} &= \dfrac{7}{12}\left(5\expval*{g^2}-1\right) = \num{-0.192+-0.019},
        &
        d_4^{(3/2)} &= \dfrac{77}{8}\expval*{g^4}-\dfrac{77}{12}\expval*{g^2}+\dfrac{11}{24} = \num{0.007+-0.006}.
    \end{aligned}
\end{equation}
\begin{table}
    \centering
    \input{tables/da-gegenbauer-moments-coefficients}
    \caption{First nonzero coefficients $A_n$ and $B_{n,j}$ used to relate the moments $\expval*{g^n}$ and the Gegenbauer moments $d_n^{(3/2)}$ of the \gls{da}.}
    \label{tab:da-gegenbauer-moments-coefficients}
\end{table}

%% file: tables/da-gegenbauer-moments-coefficients.tex
\begin{tabular}{cc ccc}
\toprule
$n$   & 0 & 2     & 4      & 6      \\
$A_n$ & 1 & $1/5$ & $3/35$ & $1/21$ \\
\end{tabular}
\newline
\begin{tabular}{c ccc ccc}
\midrule
$n$       & 1     & 2       & 3      & 3      & 4      & 4      \\
$j$       & 1     & 2       & 1      & 3      & 2      & 4      \\
$B_{n,j}$ & $3/5$ & $12/35$ & $9/35$ & $4/21$ & $8/35$ & $8/77$ \\
\bottomrule
\end{tabular}

%% file: appendices/nrqcd.tex
\section{NRQCD approach}
\label{sec:nrqcd}

In this appendix, we gather the equations used to compute the \gls{nrqcd} prediction of the \gls{da}, which is given in $x$-space as (see \cite{Chung:2019ota} and references therein)
\begin{equation}
    \label{eq:nrqcd-da}
    \phi(x,\mu)
    =
    ~ \phi^{(0)}(x)
    + \dfrac{\alpha_s(\mu)C_F}{4\pi} \phi^{(1)}(x)
    + \expval*{v^2}\phi^{(v^2)}(x)
    + \order{\alpha_s^2, \alpha_s v, v^4}.
\end{equation}
The parameter $\expval*{v^2}=\num{0.222(70)}$ \cite{Chung:2019ota} is the relativistic correction to the   \gls{ldme}. The other functions are given by \cite{Wang:2013ywc,Bodwin:2014bpa,Wang:2017bgv}
\begin{equation}
    \begin{aligned}
        \phi^{(0)}(x) &=\delta(x-1/2),\\
        \phi^{(1)}(x) &=
        \left(\log\dfrac{\mu_0^2}{m_Q^2}-1\right) \left[4x\dfrac{1+2\bar{x}}{1-2x}\theta(1-2x)\right]_{++}
        -\left[8x\dfrac{1+2\bar{x}}{1-2x}\log(1-2x)\theta(1-2x)\right]_{++} \\
        &+\left[\dfrac{16x\bar{x}}{(1-2x)^2}\theta(1-2x)\right]_{+++}
        + \Delta\left[16x\theta(1-2x)\right]_{++}
        + (x \longleftrightarrow \bar{x}),\\
        \phi^{(v^2)}(x) &= \dfrac{1}{24} \delta^{(2)}\left(x-\dfrac{1}{2}\right).
    \end{aligned}
\end{equation}
The plus prescriptions are given by
\begin{equation}
    \begin{aligned}
        \int_0^1 \dd{x}[f(x)]_{++} g(x) &= \int_0^1 \dd{x} f(x)(g(x)-g(1/2)), \\
        \int_0^1 \dd{x}[f(x)]_{+++} g(x) &= \int_0^1 \dd{x} f(x)(g(x)-g(1/2)-g^{\prime}(1/2)(x-1/2)),
    \end{aligned}
\end{equation}
for generic functions $f(x)$ and $g(x)$, and the \gls{itd} of \cref{eq:nrqcd-da} is
\begin{equation}
    \tilde{\phi}(\nu,\mu) = 1+\dfrac{\alpha_s(\mu)C_F}{4\pi}\left(\left[\log\left[\dfrac{\mu_0^2}{m_Q^2}\right]-1\right]A(\nu)+B(\nu)+C(\nu)+\Delta D(\nu)\right)-\expval*{v^2}\dfrac{\nu^2}{24},
\end{equation}
where the quantum corrections are
\begin{equation}
    \begin{aligned}
        A(\nu) &=4\Ci\left(\frac{\nu}{2}\right)-4\log(\frac{\nu}{2})-4\gamma-\dfrac{8}{\nu}\sin(\frac{\nu}{2})
        +3+\dfrac{8}{\nu^2}\left(1-\cos(\frac{\nu}{2})\right), \\
        B(\nu) &=\dfrac{\nu^2}{4} {}_{3}\mathrm{F}_4\left(1,1,1,\frac{3}{2},2,2,2,-\frac{\nu^2}{16}\right)
        -4 {}_{2}\mathrm{F}_3\left(\frac{1}{2},\frac{1}{2},\frac{1}{2},\frac{3}{2},\frac{3}{2},-\frac{\nu^2}{16}\right) \\
        &-{}_{2}\mathrm{F}_3\left(1,1,\frac{1}{2},2,2,-\frac{\nu^2}{16}\right)+5, \\
        C(\nu) &=8-4\cos(\frac{\nu}{2})-\frac{8}{\nu}\sin(\frac{\nu}{2})-2\nu\Si\left(\frac{\nu}{2}\right), \\
        D(\nu) &=4+\frac{32}{\nu^2}\left(1-\cos(\frac{\nu}{2})\right).
    \end{aligned}
\end{equation}
$\Ci$ and $\Si$ are the cosine and sine integrals and ${}_pF_q$ are hypergeometric functions. Since $p \leq q$, their convergence is guaranteed for all Ioffe times \cite{196938}. The parenthesis multiplying $\alpha_s$ vanishes at $\nu=0$, such that $\tilde{\phi}(\nu=0,\mu)=1$. To plot the  \gls{da} we need several parameters: we take the typical energy scale for the \gls{nrqcd} \gls{ldme}, $\mu_0$, to be $2\bar{m}_c$, the $\msbar$ charm-quark mass $\bar{m}_c = \qty{1.27(2)}{\giga\eV}$ \cite{Workman:2022ynf}, and we take the one-loop corrected $\msbar$ mass for a given quark \cite{TARRACH1981384}
\begin{equation}
    m_c=\bar{m}_c\left(1+\dfrac{1}{\pi}\alpha(\bar{m}_c)C_F\right)=\qty{1.46(2)}{\giga\eV}.
\end{equation}
The even Mellin moments of \cref{eq:nrqcd-da} are given by
\begin{equation}
    \begin{aligned}
        \expval{g^{2n}} &=
        \dfrac{\expval*{v^2}}{3} \delta_{n,1}
        + \dfrac{\alpha_s(\mu)C_F}{2\pi}
        \Bigg\{
          \left[\log\left(\frac{\mu_0^2}{m_Q^2}\right)-1\right] \left(\dfrac{1}{n}-\frac{1}{2n+1}-\dfrac{1}{2n+2}\right)
          \\
          &+\frac{1}{n^2}-\frac{2}{(2n+1)^2}-\frac{1}{2(1+n)^2}+\frac{2}{2n-1}-\frac{2}{2n+1}
          \\
          &+4\Delta\left(\frac{1}{2n+1}-\frac{1}{2n+2}\right)
        \Bigg\}.
    \end{aligned}
\end{equation}

%% file: appendices/ds.tex
\section{Dyson-Schwinger approach}
\label{sec:ds}

In this section, we gather the equations used in the \gls{ds} approach to compute the \gls{da}.
The authors of \cite{Ding:2015rkn} propose to model the \gls{da} like a finite-width representation of the static limit, $\delta(x-1/2)$. That is, they propose a function that is convex-concave-convex in the range $x\in[0,1]$,
\begin{equation}
    \label{eq:ds-da-x-space}
    \phi(g) = \dfrac{3}{2} N_{\lambda} (1-g^2) e^{-(\lambda g)^2}
\end{equation}
where $\lambda=\num{1.70(27)}$. Denoting the error function by $\erf$, the normalization is
\begin{equation}
N_\lambda^{-1} = \dfrac{1}{8\lambda^3} \left(3(2\lambda^2-1)\sqrt{\pi}\erf(\lambda)+6\lambda e^{-\lambda^2}\right).
\end{equation}
The \gls{itd} corresponding to \cref{eq:ds-da-x-space} is
\begin{equation}
\begin{aligned}
\phi(\nu,\mu) &= \dfrac{3}{4\lambda} N_{\lambda}
\Bigg[
  \sqrt{\pi} \left(1-\dfrac{1}{2\lambda^2}+\dfrac{\nu^2}{16\lambda^4}\right)\exp(-\dfrac{\nu^2}{16\lambda^2})
  \real \erf\left(\lambda-\dfrac{\nu\iu}{4\lambda}\right)
  \\
  &+ \left(\dfrac{1}{\lambda}\cos\left(\dfrac{\nu}{2}\right)-\dfrac{\nu}{4\lambda^3}\sin\left(\dfrac{\nu}{2}\right)\right)
  e^{-\lambda^2}
\Bigg].
\end{aligned}
\end{equation}
The even moments of \cref{eq:ds-da-x-space} are
\begin{equation}
\expval{g^{2n}}
=
\dfrac{3}{4} \dfrac{N_{\lambda}}{\lambda^{2n+1}} \left[\gamma\left(n+1/2,\lambda^2\right)-\dfrac{1}{\lambda^2}\gamma\left(n+3/2,\lambda^2\right)\right]
\end{equation}
where $\gamma$ is the lower incomplete gamma function. The odd moments are zero, while $\expval{1}=1$ so that the \gls{da} is normalized.

%% file: main.bbl
\providecommand{\href}[2]{#2}\begingroup\raggedright\begin{thebibliography}{10}

\bibitem{Collins:1989gx}
J.C.~Collins, D.E.~Soper and G.F.~Sterman, \emph{{Factorization of Hard
  Processes in QCD}},
  \href{https://doi.org/10.1142/9789814503266_0001}{\emph{Adv. Ser. Direct.
  High Energy Phys.} {\bfseries 5} (1989) 1}
  [\href{https://arxiv.org/abs/hep-ph/0409313}{{\ttfamily hep-ph/0409313}}].

\bibitem{Radyushkin:1996ru}
A.V.~Radyushkin, \emph{{Asymmetric gluon distributions and hard diffractive
  electroproduction}},
  \href{https://doi.org/10.1016/0370-2693(96)00844-1}{\emph{Phys. Lett. B}
  {\bfseries 385} (1996) 333}
  [\href{https://arxiv.org/abs/hep-ph/9605431}{{\ttfamily hep-ph/9605431}}].

\bibitem{Collins:1996fb}
J.C.~Collins, L.~Frakfurt and M.~Strikman, \emph{{Factorization for hard
  exclusive electroproduction of mesons in QCD}},
  \href{https://doi.org/10.1103/PhysRevD.56.2982}{\emph{Phys. Rev. D}
  {\bfseries 56} (1997) 2982}
  [\href{https://arxiv.org/abs/hep-ph/9611433}{{\ttfamily hep-ph/9611433}}].

\bibitem{Lepage:1980fj}
G.P.~Lepage and S.J.~Brodsky, \emph{{Exclusive Processes in Perturbative
  Quantum Chromodynamics}},
  \href{https://doi.org/10.1103/PhysRevD.22.2157}{\emph{Phys. Rev. D}
  {\bfseries 22} (1980) 2157}.

\bibitem{Diehl:2003ny}
M.~Diehl, \emph{{Generalized parton distributions}},
  \href{https://doi.org/10.1016/j.physrep.2003.08.002}{\emph{Phys. Rept.}
  {\bfseries 388} (2003) 41}
  [\href{https://arxiv.org/abs/hep-ph/0307382}{{\ttfamily hep-ph/0307382}}].

\bibitem{Bodwin:1994jh}
G.T.~Bodwin, E.~Braaten and G.P.~Lepage, \emph{{Rigorous QCD analysis of
  inclusive annihilation and production of heavy quarkonium}},
  \href{https://doi.org/10.1103/PhysRevD.55.5853}{\emph{Phys. Rev. D}
  {\bfseries 51} (1995) 1125}
  [\href{https://arxiv.org/abs/hep-ph/9407339}{{\ttfamily hep-ph/9407339}}].

\bibitem{Chung:2019ota}
Chung, Ee, Kang, Kim, Lee and Wang, \emph{{Pseudoscalar Quarkonium+gamma
  Production at NLL+NLO accuracy}},
  \href{https://doi.org/10.1007/JHEP10(2019)162}{\emph{JHEP} {\bfseries 10}
  (2019) 162} [\href{https://arxiv.org/abs/1906.03275}{{\ttfamily
  1906.03275}}].

\bibitem{Ding:2015rkn}
M.~Ding, F.~Gao, L.~Chang, Y.-X.~Liu and C.D.~Roberts, \emph{{Leading-twist
  parton distribution amplitudes of S-wave heavy-quarkonia}},
  \href{https://doi.org/10.1016/j.physletb.2015.11.075}{\emph{Phys. Lett. B}
  {\bfseries 753} (2016) 330}
  [\href{https://arxiv.org/abs/1511.04943}{{\ttfamily 1511.04943}}].

\bibitem{Binosi:2018rht}
D.~Binosi, L.~Chang, M.~Ding, F.~Gao, J.~Papavassiliou and C.D.~Roberts,
  \emph{{Distribution Amplitudes of Heavy-Light Mesons}},
  \href{https://doi.org/10.1016/j.physletb.2019.01.033}{\emph{Phys. Lett. B}
  {\bfseries 790} (2019) 257}
  [\href{https://arxiv.org/abs/1812.05112}{{\ttfamily 1812.05112}}].

\bibitem{Serna:2020txe}
F.E.~Serna, R.C.~da~Silveira, J.J.~Cobos-Mart\'\i{}nez, B.~El-Bennich and
  E.~Rojas, \emph{{Distribution amplitudes of heavy mesons and quarkonia on the
  light front}},
  \href{https://doi.org/10.1140/epjc/s10052-020-08517-3}{\emph{Eur. Phys. J. C}
  {\bfseries 80} (2020) 955}
  [\href{https://arxiv.org/abs/2008.09619}{{\ttfamily 2008.09619}}].

\bibitem{Arifi:2024mff}
A.J.~Arifi, L.~Happ, S.~Ohno and M.~Oka, \emph{{Structure of Heavy Mesons in
  the Light-Front Quark Model}},
  \href{https://arxiv.org/abs/2401.07933}{{\ttfamily 2401.07933}}.

\bibitem{Braun:1997kw}
V.M.~Braun, \emph{{Light cone sum rules}},  in \emph{{4th International
  Workshop on Progress in Heavy Quark Physics}}, pp.~105--118, 9, 1997
  [\href{https://arxiv.org/abs/hep-ph/9801222}{{\ttfamily hep-ph/9801222}}].

\bibitem{Braun:2006dg}
V.M.~Braun et~al., \emph{{Moments of pseudoscalar meson distribution amplitudes
  from the lattice}},
  \href{https://doi.org/10.1103/PhysRevD.74.074501}{\emph{Phys. Rev. D}
  {\bfseries 74} (2006) 074501}
  [\href{https://arxiv.org/abs/hep-lat/0606012}{{\ttfamily hep-lat/0606012}}].

\bibitem{Braguta:2006wr}
V.V.~Braguta, A.K.~Likhoded and A.V.~Luchinsky, \emph{{The Study of leading
  twist light cone wave function of eta(c) meson}},
  \href{https://doi.org/10.1016/j.physletb.2007.01.014}{\emph{Phys. Lett. B}
  {\bfseries 646} (2007) 80}
  [\href{https://arxiv.org/abs/hep-ph/0611021}{{\ttfamily hep-ph/0611021}}].

\bibitem{Braguta:2007fh}
V.V.~Braguta, \emph{{The study of leading twist light cone wave functions of
  J/psi meson}}, \href{https://doi.org/10.1103/PhysRevD.75.094016}{\emph{Phys.
  Rev. D} {\bfseries 75} (2007) 094016}
  [\href{https://arxiv.org/abs/hep-ph/0701234}{{\ttfamily hep-ph/0701234}}].

\bibitem{Braguta:2007tq}
V.V.~Braguta, \emph{{The Study of leading twist light cone wave functions of 2S
  state charmonium mesons}},
  \href{https://doi.org/10.1103/PhysRevD.77.034026}{\emph{Phys. Rev. D}
  {\bfseries 77} (2008) 034026}
  [\href{https://arxiv.org/abs/0709.3885}{{\ttfamily 0709.3885}}].

\bibitem{Braguta:2008qe}
V.V.~Braguta, A.K.~Likhoded and A.V.~Luchinsky, \emph{{Leading twist
  distribution amplitudes of P-wave nonrelativistic mesons}},
  \href{https://doi.org/10.1103/PhysRevD.79.074004}{\emph{Phys. Rev. D}
  {\bfseries 79} (2009) 074004}
  [\href{https://arxiv.org/abs/0810.3607}{{\ttfamily 0810.3607}}].

\bibitem{RQCD:2019osh}
{\scshape RQCD} collaboration, \emph{{Light-cone distribution amplitudes of
  pseudoscalar mesons from lattice QCD}},
  \href{https://doi.org/10.1007/JHEP08(2019)065}{\emph{JHEP} {\bfseries 08}
  (2019) 065} [\href{https://arxiv.org/abs/1903.08038}{{\ttfamily
  1903.08038}}].

\bibitem{RQCD:2019hps}
{\scshape RQCD} collaboration, \emph{{Light-cone distribution amplitudes of
  octet baryons from lattice QCD}},
  \href{https://doi.org/10.1140/epja/i2019-12803-6}{\emph{Eur. Phys. J. A}
  {\bfseries 55} (2019) 116}
  [\href{https://arxiv.org/abs/1903.12590}{{\ttfamily 1903.12590}}].

\bibitem{Ji:2013dva}
X.~Ji, \emph{{Parton Physics on a Euclidean Lattice}},
  \href{https://doi.org/10.1103/PhysRevLett.110.262002}{\emph{Phys. Rev. Lett.}
  {\bfseries 110} (2013) 262002}
  [\href{https://arxiv.org/abs/1305.1539}{{\ttfamily 1305.1539}}].

\bibitem{Alexandrou:2019lfo}
C.~Alexandrou, K.~Cichy, M.~Constantinou, K.~Hadjiyiannakou, K.~Jansen,
  A.~Scapellato et~al., \emph{{Systematic uncertainties in parton distribution
  functions from lattice QCD simulations at the physical point}},
  \href{https://doi.org/10.1103/PhysRevD.99.114504}{\emph{Phys. Rev. D}
  {\bfseries 99} (2019) 114504}
  [\href{https://arxiv.org/abs/1902.00587}{{\ttfamily 1902.00587}}].

\bibitem{Karpie:2021pap}
{\scshape HadStruc} collaboration, \emph{{The continuum and leading twist
  limits of parton distribution functions in lattice QCD}},
  \href{https://doi.org/10.1007/JHEP11(2021)024}{\emph{JHEP} {\bfseries 11}
  (2021) 024} [\href{https://arxiv.org/abs/2105.13313}{{\ttfamily
  2105.13313}}].

\bibitem{Joo:2019bzr}
B.~Jo\'o, J.~Karpie, K.~Orginos, A.V.~Radyushkin, D.G.~Richards, R.S.~Sufian
  et~al., \emph{{Pion valence structure from Ioffe-time parton
  pseudodistribution functions}},
  \href{https://doi.org/10.1103/PhysRevD.100.114512}{\emph{Phys. Rev. D}
  {\bfseries 100} (2019) 114512}
  [\href{https://arxiv.org/abs/1909.08517}{{\ttfamily 1909.08517}}].

\bibitem{Gao:2022vyh}
X.~Gao, A.D.~Hanlon, N.~Karthik, S.~Mukherjee, P.~Petreczky, P.~Scior et~al.,
  \emph{{Pion distribution amplitude at the physical point using the
  leading-twist expansion of the quasi-distribution-amplitude matrix element}},
  \href{https://doi.org/10.1103/PhysRevD.106.074505}{\emph{Phys. Rev. D}
  {\bfseries 106} (2022) 074505}
  [\href{https://arxiv.org/abs/2206.04084}{{\ttfamily 2206.04084}}].

\bibitem{LatticeParton:2022zqc}
{\scshape Lattice Parton} collaboration, \emph{{Pion and Kaon Distribution
  Amplitudes from Lattice QCD}},
  \href{https://doi.org/10.1103/PhysRevLett.129.132001}{\emph{Phys. Rev. Lett.}
  {\bfseries 129} (2022) 132001}
  [\href{https://arxiv.org/abs/2201.09173}{{\ttfamily 2201.09173}}].

\bibitem{Baker:2024zcd}
E.~Baker, D.~Bollweg, P.~Boyle, I.~Clo\"et, X.~Gao, S.~Mukherjee et~al.,
  \emph{{Lattice QCD calculation of the pion distribution amplitude with domain
  wall fermions at physical pion mass}},
  \href{https://arxiv.org/abs/2405.20120}{{\ttfamily 2405.20120}}.

\bibitem{Kovner:2024pwl}
{\scshape HadStruc} collaboration, \emph{{Extracting the Pion Distribution
  Amplitude from Lattice QCD through Pseudo-Distributions}},
  \href{https://doi.org/10.22323/1.453.0300}{\emph{PoS} {\bfseries LATTICE2023}
  (2024) 300} [\href{https://arxiv.org/abs/2401.06858}{{\ttfamily
  2401.06858}}].

\bibitem{Holligan:2024umc}
J.~Holligan and H.-W.~Lin, \emph{{Pion valence quark distribution at physical
  pion mass of N $_{f}$ = 2 + 1 + 1 lattice QCD}},
  \href{https://doi.org/10.1088/1361-6471/ad3162}{\emph{J. Phys. G} {\bfseries
  51} (2024) 065101} [\href{https://arxiv.org/abs/2404.14525}{{\ttfamily
  2404.14525}}].

\bibitem{Bhattacharya:2024qpp}
S.~Bhattacharya, K.~Cichy, M.~Constantinou, A.~Metz, N.~Nurminen and
  F.~Steffens, \emph{{Generalized parton distributions from the
  pseudo-distribution approach on the lattice}},
  \href{https://arxiv.org/abs/2405.04414}{{\ttfamily 2405.04414}}.

\bibitem{Dutrieux:2024umu}
H.~Dutrieux, R.~Edwards, C.~Egerer, J.~Karpie, C.~Monahan, K.~Orginos et~al.,
  \emph{{Towards Unpolarized GPDs from Pseudo-Distributions}},
  \href{https://arxiv.org/abs/2405.10304}{{\ttfamily 2405.10304}}.

\bibitem{Zhao:2020bsx}
S.~Zhao and A.V.~Radyushkin, \emph{{$B$-meson Ioffe-time distribution amplitude
  at short distances}},
  \href{https://doi.org/10.1103/PhysRevD.103.054022}{\emph{Phys. Rev. D}
  {\bfseries 103} (2021) 054022}
  [\href{https://arxiv.org/abs/2006.05663}{{\ttfamily 2006.05663}}].

\bibitem{Constantinou:2020pek}
M.~Constantinou, \emph{{The x-dependence of hadronic parton distributions: A
  review on the progress of lattice QCD}},
  \href{https://doi.org/10.1140/epja/s10050-021-00353-7}{\emph{Eur. Phys. J. A}
  {\bfseries 57} (2021) 77} [\href{https://arxiv.org/abs/2010.02445}{{\ttfamily
  2010.02445}}].

\bibitem{Radyushkin:1977gp}
A.V.~Radyushkin, \emph{{Deep Elastic Processes of Composite Particles in Field
  Theory and Asymptotic Freedom}},
  \href{https://arxiv.org/abs/hep-ph/0410276}{{\ttfamily hep-ph/0410276}}.

\bibitem{Radyushkin:2019owq}
A.V.~Radyushkin, \emph{{Generalized parton distributions and
  pseudodistributions}},
  \href{https://doi.org/10.1103/PhysRevD.100.116011}{\emph{Phys. Rev. D}
  {\bfseries 100} (2019) 116011}
  [\href{https://arxiv.org/abs/1909.08474}{{\ttfamily 1909.08474}}].

\bibitem{Radyushkin:2017cyf}
A.V.~Radyushkin, \emph{{Quasi-parton distribution functions, momentum
  distributions, and pseudo-parton distribution functions}},
  \href{https://doi.org/10.1103/PhysRevD.96.034025}{\emph{Phys. Rev. D}
  {\bfseries 96} (2017) 034025}
  [\href{https://arxiv.org/abs/1705.01488}{{\ttfamily 1705.01488}}].

\bibitem{Ioffe:1969kf}
B.L.~Ioffe, \emph{{Space-time picture of photon and neutrino scattering and
  electroproduction cross-section asymptotics}},
  \href{https://doi.org/10.1016/0370-2693(69)90415-8}{\emph{Phys. Lett. B}
  {\bfseries 30} (1969) 123}.

\bibitem{Radyushkin:2016hsy}
A.~Radyushkin, \emph{{Nonperturbative Evolution of Parton
  Quasi-Distributions}},
  \href{https://doi.org/10.1016/j.physletb.2017.02.019}{\emph{Phys. Lett. B}
  {\bfseries 767} (2017) 314}
  [\href{https://arxiv.org/abs/1612.05170}{{\ttfamily 1612.05170}}].

\bibitem{Belitsky:2005qn}
A.V.~Belitsky and A.V.~Radyushkin, \emph{{Unraveling hadron structure with
  generalized parton distributions}},
  \href{https://doi.org/10.1016/j.physrep.2005.06.002}{\emph{Phys. Rept.}
  {\bfseries 418} (2005) 1}
  [\href{https://arxiv.org/abs/hep-ph/0504030}{{\ttfamily hep-ph/0504030}}].

\bibitem{Orginos:2017kos}
K.~Orginos, A.~Radyushkin, J.~Karpie and S.~Zafeiropoulos, \emph{{Lattice QCD
  exploration of parton pseudo-distribution functions}},
  \href{https://doi.org/10.1103/PhysRevD.96.094503}{\emph{Phys. Rev. D}
  {\bfseries 96} (2017) 094503}
  [\href{https://arxiv.org/abs/1706.05373}{{\ttfamily 1706.05373}}].

\bibitem{Karpie:2018zaz}
J.~Karpie, K.~Orginos and S.~Zafeiropoulos, \emph{{Moments of Ioffe time parton
  distribution functions from non-local matrix elements}},
  \href{https://doi.org/10.1007/JHEP11(2018)178}{\emph{JHEP} {\bfseries 11}
  (2018) 178} [\href{https://arxiv.org/abs/1807.10933}{{\ttfamily
  1807.10933}}].

\bibitem{Ishikawa:2017faj}
T.~Ishikawa, Y.-Q.~Ma, J.-W.~Qiu and S.~Yoshida, \emph{{Renormalizability of
  quasiparton distribution functions}},
  \href{https://doi.org/10.1103/PhysRevD.96.094019}{\emph{Phys. Rev. D}
  {\bfseries 96} (2017) 094019}
  [\href{https://arxiv.org/abs/1707.03107}{{\ttfamily 1707.03107}}].

\bibitem{Radyushkin:2017lvu}
A.V.~Radyushkin, \emph{{Quark pseudodistributions at short distances}},
  \href{https://doi.org/10.1016/j.physletb.2018.04.023}{\emph{Phys. Lett. B}
  {\bfseries 781} (2018) 433}
  [\href{https://arxiv.org/abs/1710.08813}{{\ttfamily 1710.08813}}].

\bibitem{NIST:DLMF}
``\textit{NIST Digital Library of Mathematical Functions}.''
  \url{https://dlmf.nist.gov/}, Release 1.1.10 of 2023-06-15.

\bibitem{Braun:2003rp}
V.M.~Braun, G.P.~Korchemsky and D.~M\"uller, \emph{{The Uses of conformal
  symmetry in QCD}},
  \href{https://doi.org/10.1016/S0146-6410(03)90004-4}{\emph{Prog. Part. Nucl.
  Phys.} {\bfseries 51} (2003) 311}
  [\href{https://arxiv.org/abs/hep-ph/0306057}{{\ttfamily hep-ph/0306057}}].

\bibitem{Efremov:1979qk}
A.V.~Efremov and A.V.~Radyushkin, \emph{{Factorization and Asymptotical
  Behavior of Pion Form-Factor in QCD}},
  \href{https://doi.org/10.1016/0370-2693(80)90869-2}{\emph{Phys. Lett. B}
  {\bfseries 94} (1980) 245}.

\bibitem{Lepage:1979zb}
G.P.~Lepage and S.J.~Brodsky, \emph{{Exclusive Processes in Quantum
  Chromodynamics: Evolution Equations for Hadronic Wave Functions and the
  Form-Factors of Mesons}},
  \href{https://doi.org/10.1016/0370-2693(79)90554-9}{\emph{Phys. Lett. B}
  {\bfseries 87} (1979) 359}.

\bibitem{Workman:2022ynf}
{\scshape Particle Data Group} collaboration, \emph{{Review of Particle
  Physics}}, \href{https://doi.org/10.1093/ptep/ptac097}{\emph{PTEP} {\bfseries
  2022} (2022) 083C01}.

\bibitem{FlavourLatticeAveragingGroupFLAG:2021npn}
{\scshape Flavour Lattice Averaging Group (FLAG)} collaboration, \emph{{FLAG
  Review 2021}},
  \href{https://doi.org/10.1140/epjc/s10052-022-10536-1}{\emph{Eur. Phys. J. C}
  {\bfseries 82} (2022) 869}
  [\href{https://arxiv.org/abs/2111.09849}{{\ttfamily 2111.09849}}].

\bibitem{196938}
\emph{Chapter iii hypergeometric functions},  in \emph{The Special Functions
  and Their Approximations}, Y.L.~Luke, ed., vol.~53 of \emph{Mathematics in
  Science and Engineering}, pp.~38--114, Elsevier (1969),
  \href{https://doi.org/https://doi.org/10.1016/S0076-5392(08)62627-2}{DOI}.

\bibitem{Fritzsch:2012wq}
P.~Fritzsch, F.~Knechtli, B.~Leder, M.~Marinkovic, S.~Schaefer, R.~Sommer
  et~al., \emph{{The strange quark mass and Lambda parameter of two flavor
  QCD}}, \href{https://doi.org/10.1016/j.nuclphysb.2012.07.026}{\emph{Nucl.
  Phys. B} {\bfseries 865} (2012) 397}
  [\href{https://arxiv.org/abs/1205.5380}{{\ttfamily 1205.5380}}].

\bibitem{Heitger:2013oaa}
J.~Heitger, G.M.~von Hippel, S.~Schaefer and F.~Virotta, \emph{{Charm quark
  mass and D-meson decay constants from two-flavour lattice QCD}},
  \href{https://doi.org/10.22323/1.187.0475}{\emph{PoS} {\bfseries LATTICE2013}
  (2014) 475} [\href{https://arxiv.org/abs/1312.7693}{{\ttfamily 1312.7693}}].

\bibitem{openqcd20}
M.~Luscher and A.~Schafer, ``{OpenQCD: Simulation programs for lattice QCD}.''

\bibitem{Luscher:2003qa}
M.~Luscher, \emph{{Solution of the Dirac equation in lattice QCD using a domain
  decomposition method}},
  \href{https://doi.org/10.1016/S0010-4655(03)00486-7}{\emph{Comput. Phys.
  Commun.} {\bfseries 156} (2004) 209}
  [\href{https://arxiv.org/abs/hep-lat/0310048}{{\ttfamily hep-lat/0310048}}].

\bibitem{Luscher:2007es}
M.~Luscher, \emph{{Deflation acceleration of lattice QCD simulations}},
  \href{https://doi.org/10.1088/1126-6708/2007/12/011}{\emph{JHEP} {\bfseries
  12} (2007) 011} [\href{https://arxiv.org/abs/0710.5417}{{\ttfamily
  0710.5417}}].

\bibitem{Luscher:2007se}
M.~Luscher, \emph{{Local coherence and deflation of the low quark modes in
  lattice QCD}},
  \href{https://doi.org/10.1088/1126-6708/2007/07/081}{\emph{JHEP} {\bfseries
  07} (2007) 081} [\href{https://arxiv.org/abs/0706.2298}{{\ttfamily
  0706.2298}}].

\bibitem{ddhmc}
M.~Luscher and A.~Scafer, ``{DD-HMC: Simulation program for two-flavour lattice
  QCD}.''

\bibitem{Wolff:2003sm}
{\scshape ALPHA} collaboration, \emph{{Monte Carlo errors with less errors}},
  \href{https://doi.org/10.1016/S0010-4655(03)00467-3}{\emph{Comput. Phys.
  Commun.} {\bfseries 156} (2004) 143}
  [\href{https://arxiv.org/abs/hep-lat/0306017}{{\ttfamily hep-lat/0306017}}].

\bibitem{Schaefer:2010hu}
{\scshape ALPHA} collaboration, \emph{{Critical slowing down and error analysis
  in lattice QCD simulations}},
  \href{https://doi.org/10.1016/j.nuclphysb.2010.11.020}{\emph{Nucl. Phys. B}
  {\bfseries 845} (2011) 93} [\href{https://arxiv.org/abs/1009.5228}{{\ttfamily
  1009.5228}}].

\bibitem{Ramos:2018vgu}
A.~Ramos, \emph{{Automatic differentiation for error analysis of Monte Carlo
  data}}, \href{https://doi.org/10.1016/j.cpc.2018.12.020}{\emph{Comput. Phys.
  Commun.} {\bfseries 238} (2019) 19}
  [\href{https://arxiv.org/abs/1809.01289}{{\ttfamily 1809.01289}}].

\bibitem{Joswig:2022qfe}
F.~Joswig, S.~Kuberski, J.T.~Kuhlmann and J.~Neuendorf, \emph{{pyerrors: A
  python framework for error analysis of Monte Carlo data}},
  \href{https://doi.org/10.1016/j.cpc.2023.108750}{\emph{Comput. Phys. Commun.}
  {\bfseries 288} (2023) 108750}
  [\href{https://arxiv.org/abs/2209.14371}{{\ttfamily 2209.14371}}].

\bibitem{DellaMorte:2017dyu}
M.~Della~Morte, A.~Francis, V.~G\"ulpers, G.~Herdo\'\i{}za, G.~von Hippel,
  H.~Horch et~al., \emph{{The hadronic vacuum polarization contribution to the
  muon $g-2$ from lattice QCD}},
  \href{https://doi.org/10.1007/JHEP10(2017)020}{\emph{JHEP} {\bfseries 10}
  (2017) 020} [\href{https://arxiv.org/abs/1705.01775}{{\ttfamily
  1705.01775}}].

\bibitem{Balasubramamian:2019wgx}
R.~Balasubramamian and B.~Blossier, \emph{{Decay constant of $B_s$ and $B^*_s$
  mesons from $\mathrm{N_f}=2$ lattice QCD}},
  \href{https://doi.org/10.1140/epjc/s10052-020-7965-z}{\emph{Eur. Phys. J. C}
  {\bfseries 80} (2020) 412}
  [\href{https://arxiv.org/abs/1912.09937}{{\ttfamily 1912.09937}}].

\bibitem{Sachrajda:2004mi}
C.T.~Sachrajda and G.~Villadoro, \emph{{Twisted boundary conditions in lattice
  simulations}},
  \href{https://doi.org/10.1016/j.physletb.2005.01.033}{\emph{Phys. Lett. B}
  {\bfseries 609} (2005) 73}
  [\href{https://arxiv.org/abs/hep-lat/0411033}{{\ttfamily hep-lat/0411033}}].

\bibitem{Gusken:1989qx}
S.~Gusken, \emph{{A Study of smearing techniques for hadron correlation
  functions}}, \href{https://doi.org/10.1016/0920-5632(90)90273-W}{\emph{Nucl.
  Phys. B Proc. Suppl.} {\bfseries 17} (1990) 361}.

\bibitem{APE:1987ehd}
{\scshape APE} collaboration, \emph{{Glueball Masses and String Tension in
  Lattice QCD}},
  \href{https://doi.org/10.1016/0370-2693(87)91160-9}{\emph{Phys. Lett. B}
  {\bfseries 192} (1987) 163}.

\bibitem{Karpie:2019eiq}
J.~Karpie, K.~Orginos, A.~Rothkopf and S.~Zafeiropoulos, \emph{{Reconstructing
  parton distribution functions from Ioffe time data: from Bayesian methods to
  Neural Networks}}, \href{https://doi.org/10.1007/JHEP04(2019)057}{\emph{JHEP}
  {\bfseries 04} (2019) 057}
  [\href{https://arxiv.org/abs/1901.05408}{{\ttfamily 1901.05408}}].

\bibitem{doi:10.1137/0710036}
G.H.~Golub and V.~Pereyra, \emph{The differentiation of pseudo-inverses and
  nonlinear least squares problems whose variables separate},
  \href{https://doi.org/10.1137/0710036}{\emph{SIAM Journal on Numerical
  Analysis} {\bfseries 10} (1973) 413}
  [\href{https://arxiv.org/abs/https://doi.org/10.1137/0710036}{{\ttfamily
  https://doi.org/10.1137/0710036}}].

\bibitem{Michael:1994sz}
C.~Michael and A.~McKerrell, \emph{{Fitting correlated hadron mass spectrum
  data}}, \href{https://doi.org/10.1103/PhysRevD.51.3745}{\emph{Phys. Rev. D}
  {\bfseries 51} (1995) 3745}
  [\href{https://arxiv.org/abs/hep-lat/9412087}{{\ttfamily hep-lat/9412087}}].

\bibitem{Briceno:2018lfj}
R.A.~Brice\~no, J.V.~Guerrero, M.T.~Hansen and C.J.~Monahan,
  \emph{{Finite-volume effects due to spatially nonlocal operators}},
  \href{https://doi.org/10.1103/PhysRevD.98.014511}{\emph{Phys. Rev. D}
  {\bfseries 98} (2018) 014511}
  [\href{https://arxiv.org/abs/1805.01034}{{\ttfamily 1805.01034}}].

\bibitem{Briceno:2021jlb}
R.A.~Brice\~no and C.J.~Monahan, \emph{{A model-independent framework for
  determining finite-volume effects of spatially nonlocal operators}},
  \href{https://doi.org/10.1103/PhysRevD.103.094521}{\emph{Phys. Rev. D}
  {\bfseries 103} (2021) 094521}
  [\href{https://arxiv.org/abs/2102.01814}{{\ttfamily 2102.01814}}].

\bibitem{PhysRevD.100.074502}
H.-W.~Lin and R.~Zhang, \emph{Lattice finite-volume dependence of the nucleon
  parton distributions},
  \href{https://doi.org/10.1103/PhysRevD.100.074502}{\emph{Phys. Rev. D}
  {\bfseries 100} (2019) 074502}.

\bibitem{Joo:2019jct}
B.~Jo\'o, J.~Karpie, K.~Orginos, A.~Radyushkin, D.~Richards and
  S.~Zafeiropoulos, \emph{{Parton Distribution Functions from Ioffe time
  pseudo-distributions}},
  \href{https://doi.org/10.1007/JHEP12(2019)081}{\emph{JHEP} {\bfseries 12}
  (2019) 081} [\href{https://arxiv.org/abs/1908.09771}{{\ttfamily
  1908.09771}}].

\bibitem{Radyushkin:2019mye}
A.V.~Radyushkin, \emph{{Theory and applications of parton
  pseudodistributions}},
  \href{https://doi.org/10.1142/S0217751X20300021}{\emph{Int. J. Mod. Phys. A}
  {\bfseries 35} (2020) 2030002}
  [\href{https://arxiv.org/abs/1912.04244}{{\ttfamily 1912.04244}}].

\bibitem{Chang:2013pq}
L.~Chang, I.C.~Cloet, J.J.~Cobos-Martinez, C.D.~Roberts, S.M.~Schmidt and
  P.C.~Tandy, \emph{{Imaging dynamical chiral symmetry breaking: pion wave
  function on the light front}},
  \href{https://doi.org/10.1103/PhysRevLett.110.132001}{\emph{Phys. Rev. Lett.}
  {\bfseries 110} (2013) 132001}
  [\href{https://arxiv.org/abs/1301.0324}{{\ttfamily 1301.0324}}].

\bibitem{Gao:2014bca}
F.~Gao, L.~Chang, Y.-X.~Liu, C.D.~Roberts and S.M.~Schmidt, \emph{{Parton
  distribution amplitudes of light vector mesons}},
  \href{https://doi.org/10.1103/PhysRevD.90.014011}{\emph{Phys. Rev. D}
  {\bfseries 90} (2014) 014011}
  [\href{https://arxiv.org/abs/1405.0289}{{\ttfamily 1405.0289}}].

\bibitem{Bertone:2013vaa}
V.~Bertone, S.~Carrazza and J.~Rojo, \emph{{APFEL: A PDF Evolution Library with
  QED corrections}},
  \href{https://doi.org/10.1016/j.cpc.2014.03.007}{\emph{Comput. Phys. Commun.}
  {\bfseries 185} (2014) 1647}
  [\href{https://arxiv.org/abs/1310.1394}{{\ttfamily 1310.1394}}].

\bibitem{Bertone:2017gds}
V.~Bertone, \emph{{APFEL++: A new PDF evolution library in C++}},
  \href{https://doi.org/10.22323/1.297.0201}{\emph{PoS} {\bfseries DIS2017}
  (2018) 201} [\href{https://arxiv.org/abs/1708.00911}{{\ttfamily
  1708.00911}}].

\bibitem{math11132839}
A.~Bärligea, P.~Hochstaffl and F.~Schreier, \emph{A generalized variable
  projection algorithm for least squares problems in atmospheric remote
  sensing}, \href{https://doi.org/10.3390/math11132839}{\emph{Mathematics}
  {\bfseries 11} (2023) }.

\bibitem{oleary:2013eg}
D.P.~O’Leary and B.W.~Rust, \emph{Variable projection for nonlinear least
  squares problems},
  \href{https://doi.org/10.1007/s10589-012-9492-9}{\emph{Computational
  Optimization and Applications} {\bfseries 54} (2013) 579}.

\bibitem{golub2003separable}
G.~Golub and V.~Pereyra, \emph{Separable nonlinear least squares: the variable
  projection method and its applications}, {\emph{Inverse problems} {\bfseries
  19} (2003) R1}.

\bibitem{Wang:2013ywc}
X.-P.~Wang and D.~Yang, \emph{{The leading twist light-cone distribution
  amplitudes for the S-wave and P-wave quarkonia and their applications in
  single quarkonium exclusive productions}},
  \href{https://doi.org/10.1007/JHEP06(2014)121}{\emph{JHEP} {\bfseries 06}
  (2014) 121} [\href{https://arxiv.org/abs/1401.0122}{{\ttfamily 1401.0122}}].

\bibitem{Bodwin:2014bpa}
G.T.~Bodwin, H.S.~Chung, J.-H.~Ee, J.~Lee and F.~Petriello, \emph{{Relativistic
  corrections to Higgs boson decays to quarkonia}},
  \href{https://doi.org/10.1103/PhysRevD.90.113010}{\emph{Phys. Rev. D}
  {\bfseries 90} (2014) 113010}
  [\href{https://arxiv.org/abs/1407.6695}{{\ttfamily 1407.6695}}].

\bibitem{Wang:2017bgv}
W.~Wang, J.~Xu, D.~Yang and S.~Zhao, \emph{{Relativistic corrections to
  light-cone distribution amplitudes of S-wave B$_{c}$ mesons and heavy
  quarkonia}}, \href{https://doi.org/10.1007/JHEP12(2017)012}{\emph{JHEP}
  {\bfseries 12} (2017) 012}
  [\href{https://arxiv.org/abs/1706.06241}{{\ttfamily 1706.06241}}].

\bibitem{TARRACH1981384}
R.~Tarrach, \emph{The pole mass in perturbative qcd},
  \href{https://doi.org/https://doi.org/10.1016/0550-3213(81)90140-1}{\emph{Nuclear
  Physics B} {\bfseries 183} (1981) 384}.

\end{thebibliography}\endgroup
